\documentclass{ss_areviews}
\usepackage{ARAstroBib}
\usepackage{ar-patch}

\renewcommand{\contentsname}{\LARGE \bf TABLE OF CONTENTS\\[2ex]}

\begin{document}
\input epsf.tex    
\input epsf.def   

\input psfig.sty

\jname{Annu. Rev. Astron. \& Astrophys.}
\jyear{2006}
\jvol{44}
\ARinfo{1056-8700/97/0610-00}

\title{X-RAY EMISSION FROM EXTRAGALACTIC JETS}

\markboth{Harris \& Krawczynski}{X-ray Jets}

\author{D. E. Harris
\affiliation{Harvard-Smithsonian Center for Astrophysics}
Henric Krawczynski
\affiliation{Washington University in St. Louis}}


\begin{abstract}
This review focuses on the X-ray emission processes of extra-galactic
jets on scales resolvable by the sub arcsec resolution of the Chandra
X-ray Observatory.  It is divided into 4 parts.  The introductory
chapter reviews the classical problems for jets, as well as those
associated directly with the X-ray emission.  Throughout this section,
we deal with the dualisms of low powered radio sources versus high
powered radio galaxies and quasars; synchrotron models versus inverse
Compton models; and the distinction between the relativistic plasma
responsible for the received radiation and the medium responsible for
the transport of energy down the jet.  The second part collects the
observational and inferred parameters for the currently detected X-ray
jets and attempts to put their relative sizes and luminosities in
perspective.  In part 3, we first give the relevant radio and optical
jet characteristics, and then examine the details of the X-ray data
and how they can be related to various jet attributes.  The last
section is devoted to a critique of the two non-thermal emission
processes and to prospects for progress in our understanding of jets.
\end{abstract}

\maketitle

\section{THE PROBLEMS}\label{sec:intro}
Jets are giant collimated plasma outflows associated with some types
of Active Galactic Nuclei (AGN). The first jet was discovered in 1918
within the elliptical galaxy M87 in the Virgo cluster: ``A
curious straight ray lies in a gap in the nebulosity in
p.a. 20$^{\circ}$, apparently connected with the nucleus by a thin
line of matter.  The ray is brightest at its inner end, which is
11$^{\prime\prime}$ from the nucleus.'' \citep{curt18}.  At that time,
the extended feature was a mere curiosity and its nature was not
understood. When radio telescopes with good angular resolution and
high sensitivities became available in the sixties, it was found that
many galaxies exhibited extended radio emission consisting of a
nuclear component, jets, hotspot complexes, and radio lobes. According
to the standard picture, jets originate in the vicinity of a
super-massive black hole ('SMBH' with several million to several
billion solar masses) located at the center of the AGN;
(c.f. the early ideas of \citet{salp64}).  The jets are most likely
powered by these black holes, and the jets themselves transport
energy, momentum, and angular momentum over vast distances
\citep{rees71,blan74,sche74}, from the ``tiny'' black hole of radius
$r\,=$ $10^{-4}$ $M_{\rm BH}/10^9 M_\odot$ pc to radio hotspots,
hotspot complexes and lobes which may be a Mpc or more away.  Thus the
study of jets must address a range of scales covering a factor of
10$^{10}$!

\marginnote{{\bf CXO}
Chandra X-ray Observatory:  NASA's first X-ray imaging satellite
                with sub-arc-second resolution.  Launched in July 1999.}

Even now, after thirty years of intensive studies of radio galaxies in
the radio regime, no consensus has emerged on their fundamental
attributes such as composition, formation, and collimation.  With the
advent of the Hubble Space Telescope (HST) and the Chandra X-ray
Observatory (CXO), the optical and X-ray emission from jets can be
studied and new tests can be evaluated which were not possible based
on radio data alone.  This follows because the radio, optical, and
X-ray jet emissions are emitted by electrons with quite different
energies (i.e. Lorentz factors, $\gamma$).

\marginnote{{\bf Lorentz factor:}
for relativistic electrons, $\gamma=\frac{E}{m_e\times
c^2}$; for the jet's bulk velocity, $\beta=\frac{v}{c}$,
$\Gamma=\frac{1}{\sqrt{1-\beta^2}}$.}

This review is focused on what X-ray observations of relativistic jets
can contribute to our understanding of the physical processes in jets.
Although some jet detections were made with the imaging X-ray
observatories {\it Einstein} and {\it ROSAT}, significant progress
blossomed only with the CXO \citep{weis03} launch in 1999.  For this
reason, together with the limitations of space, we emphasize
results obtained between the years 2000 and 2005.5.  We
will concentrate on spatially resolved X-ray emission from the
kpc-scale jets. Radio observations of the pc-scale jets and broadband
observations of the spatially unresolved but highly variable core
emission from the sub-pc jets of blazar-type AGN will only be
discussed when they have direct implications for the inner workings of
kpc-jets.  Furthermore, we will not cover Galactic X-ray jets even
though they bear many similarities to their extragalactic
counterparts.  Although there have been reports of thermal X-ray
emission associated with jets (mainly in the context of 'jet-cloud
interactions'), our main concern is with the non-thermal emissions,
already well established as the major process for radio through X-ray
frequencies from multiple lines of argument including polarization
data, Faraday screen parameters, X-ray spectral fitting, and the
absence of emission lines.

Reviews on some aspects of jets include ``Theory of extragalactic
radio sources`` \citep{bege84}, ``Beams and Jets in Astrophysics''
\citep{hugh91}, ``Parsec-Scale Jets in Extragalactic Radio Sources''
\citep{zens97}, and ``Relativistic jets in AGNs'' \citep{tave04b}.
Among the many jet related meetings in the last ten years are:
``Relativistic jets in AGNs'' Cracow, 1997, \citep{ostr97}; ``Ringberg
Workshop on Relativistic Jets'' Ringberg Castle,
2001\footnote{http://www.mpa-garching.mpg.de/$\sim$ensslin/Jets/Proceedings/};
``The Physics of Relativistic Jets in the CHANDRA and XMM Era''
Bologna, 2002, \citep{brun03}; ``Triggering Relativistic Jets''
Cozumel, 2005 \citep{lee06}; and ``Ultra-Relativistic Jets in Astrophysics:
Observations, Theory and Simulations'' Banff, 2005\footnote{to
download talks:
http://www.capca.ucalgary.ca/meetings/banff2005/index.html}.
Conference reviews on the unresolved core emission have been given by
\citet{siko01,copp99,kraw04,kraw05} and \citet{tave05}.

We use the conventional definition of spectral index, $\alpha$, for
power-law radiation spectra: flux density,
S$_{\nu}~\propto~\nu^{-\alpha}$.  It is not yet known if electrons
alone, or electrons and positrons radiate the observed jet
emission. We thus refer in the following to either electrons or
electrons and positrons as ``electrons''.  We use $\gamma$ as the
Lorentz factor of particles in the jet-frame of reference, and
$\Gamma$ for the bulk Lorentz factor of the jet plasma.  As most X-ray
emitting jets are detected on only one side of otherwise double radio
sources, $\Gamma~\geq$~a few seems likely to be generally applicable.

\subsection{Jet Composition}\label{sec:1fluid}

We take the essence of a jet to be a quasi-lossless transmission line:
a conduit containing relativistically moving particles and magnetic
field (either of which could dominate the local energy) and/or
Poynting flux.  We distinguish between two substances: the ``medium''
which is responsible for delivering the power generated in the nucleus
of the host galaxy to the end of the jet and thence to the radio
lobes; and the non-thermal plasma responsible for the emission we
detect in the radio, optical, and X-ray bands. While these two
substances can be one and the same for some jet models, we prefer to
think of them as quite distinct. Most models explain the appearance of
radio, optical or X-ray bright hotspots in some jets as caused by the
transfer of some form of energy (for example, energy associated with
the medium's bulk motion or magnetic field energy), to highly
relativistic emitting particles.  The reader should note that we use
the term ``medium'' lacking more precise knowledge about the nature of
the jet material.

While the basic make-up of jets is still largely unknown, observations
of polarized radio and optical emission show that at least some of the
continuum jet emission originates as synchrotron emission from
relativistic electrons gyrating in a magnetic field. Although we have
this direct evidence about the emitting plasma, the jet medium
responsible for delivering power to the end of the jets is largely
unconstrained.  The jet medium cannot entirely consist of the
relativistic electrons that produce the observed radiation since
unavoidable inverse Compton (IC) losses off the cosmic microwave
background (CMB) photons would preclude the flow of high energy
electrons all the way to the end of some jets. Positing a minimal
magnetic field strength of 3$\mu$G, and ignoring the IC losses
associated with starlight or quasar light which would shorten the
relevant lifetimes even more, it has been shown that electrons with
$\gamma~\geq$~a few thousand cannot survive for the time required to
travel from the environs of the SMBH to the end of some jets
\citep[e.g.][]{harr06b}.


The main contenders for the underlying jet medium are Poynting flux,
electrons with $\gamma~\leq$~1000, and protons.  In addition, 'neutral
beams' have been suggested \cite[e.g. neutrons,][]{atoy04}.  The
latter hypothesis requires that the direction into which the jet is
launched changes with time to account for large scale bending and
discrete deflections such as those in 3C~120 and 3C~390.3.  Real jets
may be made of several components, or may involve the transition of a
jet dominated by one component into a jet dominated by another; e.g. a
class of models postulates that an initially electromagnetic jet
transforms into a particle dominated jet further downstream.
%
\subsection{Jet formation, structure and propagation}\label{sec:1genesis}
\subsubsection*{Jet formation}
Jets are believed to be launched from accreting supermassive black
holes and powered by either the gravitational energy of
accreting matter that moves toward the black hole or, in the
Blandford-Znajek process \citep{blan77}, by the rotational energy of a
rotating black hole. In the first case, jets may either be launched
purely electromagnetically \cite{blan76,love76}, or as the result of
magnetohydrodynamic processes at the inner regions of the accretion
disk \citep{blan82,bege84,koid99}.  In the Blandford-Znajek process,
the black hole rotating in the magnetic field supported by the
accretion disk gives rise to a Poynting flux.  Most models of jet
formation face the $\sigma$-problem ($\sigma$ is the ratio of
electromagnetic energy density to particle energy density), namely
that they predict a Poynting flux dominated energy transport by a
strongly magnetized or high-$\sigma$ plasma, while pc scale
observations indicate that the jets consist of particle dominated,
low-$\sigma$ plasma \citep{celo93,kraw02,kino02}.  Understanding the
launching of jets may thus require the solution of two problems: the
launching of a magnetically dominated outflow, and the conversion of
such an outflow into a particle dominated jet.  The latter transition
is poorly understood, and requires more theoretical work.

The process of jet formation will have an impact on the steadiness of
the jet-flow, and will affect the amplitudes and time scales of jet
luminosity variations. Modulations of the power output are believed to
cause the large amplitude brightness variations of the (unresolved)
X-ray and $\gamma$-ray emission from blazars
\citep{spad01,tani03}. Large amplitude variations on time scales of
thousands of years may be responsible for the radio, optical and X-ray
knots observed in many kpc-scale jets
\citep{staw04a,staw04b}, and the bright X-ray flare
of the M87 jet \citep{harr06a}.  Several recent studies have shown
that the flaring activity of AGN can be described in the language of
noise processes \citep{uttl05}; i.e. the study of power spectra.
Blazar flares show that the noise process that drives flares has a
rising amplitude of the power spectrum on the relatively long time
scales of a few years.  If jet knots reflect nuclear variability, it
would require substantial power at much longer time scales (red noise)
which of course are not available for direct observation.
\subsubsection*{Transverse jet structure}

\marginnote{{\bf Forward \& Reverse Shocks:} One sort of shock can
arise from the interaction of a fast medium overtaking a slower
medium.  In the frame of the contact discontinuity separating the two
media, a forward shock propagates downstream into the slower moving
medium and a reverse shock propagates upstream into the faster moving
medium.  Particle acceleration can be associated with both.  The
structure of knot C in the M87 jet (with a large gradient in
brightness, falling rapidly moving downstream) serves as an example of
a forward shock, and the expected behavior of a reverse shock is
exemplified by knot A (fig.~\ref{fig:m87}).  }
In addition to the obvious uncertainties as to the identity of the jet
medium and its bulk velocity, several jet models involve jet structure
perpendicular to the jet axis. Radio observations of transversely resolved jets
\citep[e.g.][]{swai98,lain04,lara04,push05} and theoretical models of the
core emission of blazars \citep{chia00} indicate a velocity gradient
across the jet.  Simple models use a two-zone structure, a fast moving
spine that carries most of the jet energy, surrounded by a slower
sheath, each with a characteristic value of $\Gamma$ \citep{chia00}.
\citet{lain04} assume a gradual decline of $\Gamma$ from the jet
center to the outer parts of the jet: i.e. many layers with different
velocities. If the velocity difference between layers is large, the
particles in some layers see the relativistically boosted photons from
other layers, resulting in an increase of the IC emission \citep{ghis05}.
A wealth of different jet structures has been proposed and studied in
the framework of explaining the prompt and afterglow emission from
Gamma-Ray Bursts \citep[e.g.][]{graz06} and some of these may be
relevant to kpc scale jets.

The fact that jets may have a complex structure is important for
interpreting the observational data. For example, the dominance of the
bright jet over the dim counter-jet in a number of sources was
previously thought to constrain the bulk Lorentz factor of the jets
\citep[e.g.][]{ward97}. However, the observations may merely show that
most of the radio emission comes from a slow moving plasma, and thus
may not constrain the bulk Lorentz factor of the jet component that
carries most of the jet energy and momentum.  The boundaries between
jet layers of different velocity may accelerate particles
\citep{staw02} and are of special interest for jet stability
considerations.
\subsubsection*{Jet propagation and the occurrence of knots}

The origin of jet knots (localized brightness enhancements) and the
mechanism that controls the location, strength, and longevity of the
shocks thought to be responsible for the existence of knots, have not
yet been identified unambiguously.  It is important to remember
however, that there is probably more than one type of knot and that
there are several suggested methods of producing brightness
enhancements in addition to the conventional explanation of particle
acceleration at shocks.  In the case of the M87 jet
(fig.~\ref{fig:m87}), the inner knots, D, E, and F appear quasi
regular in size and spacing suggesting a possible origin associated
with standing waves similar to those described by \citet{bere03}, or
by the elliptical mode Kelvin-Helmholtz instability \citep{loba03}.
Quite different are the knots A and C for which steep, quasi-planar
gradients in radio brightness suggest reverse and forward shocks (see
comment on shocks in the sidebar).  Note however that \citet{bick96}
have devised a detailed model of the M87 jet.  They argue that all the
knots can be explained by oblique shocks, with the apparent
differences being ascribed to relativistic effects.  Their model
requires the angle between the jet axis and the line of sight to be
30$^{\circ}$ to 35$^{\circ}$, a value substantially larger than the
10$^{\circ}$ to 20$^{\circ}$ required by the observation of fast
moving blobs downstream from the leading edge of the knot HST-1 (see
sec.~\ref{sec:3kpc}).

One of the alternative explanation of knots is that knots in
relativistic jets could be manifestations of a change in the beaming
factor.  The relativistic beaming factor, $\delta$ depends both on
$\Gamma$ and on the viewing angle, $\theta$ (the angle between the jet
axis and the line of sight in the observer's frame):

\begin{equation}
\delta^{-1}~=~\Gamma(1~-~\beta~cos~\theta).
\label{eq:delta1}
\end{equation}

\noindent
If the jet medium moves in a straight line so that $\theta$ is fixed,
an increase in $\delta$ requires a significant increase in $\Gamma$.
While we can imagine plausible ways to lower $\Gamma$, the critical
question is, are there ways to increase $\Gamma$ far from the central
engine?  This would entail a supply of energy such that the total
power flow could decrease yet $\Gamma$ could increase (e.g. by
converting some power from the flow as in magnetic reconnection).
\citet{siko05} discuss this scenario, but deal only with the situation
close to the black hole.  While there is circumstantial evidence for
acceleration of jet features on pc-scales \citep[e.g.][]{hardee05} and
it is generally accepted that both FRI \citep{lain02b} and quasar
\citep{ward97} jets decelerate on pc to kpc scales there is no
indication that significant jet acceleration occurs on kpc scales
which may be required for some IC models of X-ray emission.

\marginnote{{\bf Fanaroff-Riley class:}
FRI radio galaxies are of lower radio luminosity
than FRII's and quasars, and the brighter radio structures are close
to the nucleus.
}

If the jet medium is allowed to significantly change its direction,
modest changes in $\theta$ can produce large changes in $\delta$.
On VLBI scales, there has been a long standing debate on ballistic vs.
curved trajectories.  On the kpc scale the question arises: does the
medium move in a straight or gently curved path, or might it follow a
helical pattern controlled by a field structure of the same topology?
If the latter case holds, the changes in brightness along the jet
could be explained by beaming effects and some of the problems for
high $\Gamma$ jet models, such as excessive jet length, would be
mitigated.  \citet{bahc95} remark on the apparent helical morphology
of the HST image of 3C~273 (fig.~\ref{fig:hst273}) and \citet{naka01}
argue for a 'torsional Alv\`{e}n wave train' moving out to large
distances from the central engine as a method of controlling the large
scale structure.  There are of course numerous examples of large scale
bending \citep[e.g. 3C~120:][]{walk87} and discrete deflections
\citep[e.g. 3C~390.3][]{harr99}; but in these cases we would
anticipate deceleration only.

\marginnote{{\bf VLBI:}
Very Long Baseline Interferometry: The technique of aperture
synthesis in which the component radio telescopes are not physically
connected, thereby permitting the use of inter-continental baselines
resulting in synthesized beam sizes of milli arcsecs.
}

\begin{figure}[t]*
\caption{An HST image of the jet in 3C273.}
\label{fig:hst273}
\end{figure}

\subsubsection*{Terminal Hotspots}\label{sec:1hotspots}

Terminal hotspots, like knots, are thought to be localized volumes of
high emissivity which are produced by strong shocks or a system of
shocks.  The somewhat hazy distinction between hotspots and knots is
that downstream from a knot, the jet usually propagates much as
before, whereas at the terminal hotspot, the jet itself terminates and
the remaining flow is thought to create the radio lobes or tails.
Thus the underlying jet medium must suffer severe deceleration and the
outward flow from the hotspot is non-relativistic and is not confined
to a small angle.  This is patently not true for the so called
'primary' hotspots in double or multiple systems.  Instead of a
terminal shock, primaries (and also aberrations such as hotspot B in 3C
390.3 North) may have oblique reflectors in essence, although the
actual mechanism for bending might be more akin to refraction.
For an extensive discussion of the differences between knots and
hotspots, see \citet{brid94}.

Knots are a common property of FRI jets and generally do not
\marginnote{{\bf Two-Zone Models} In many areas of jet modeling, it
is often the case that a simple, single power law or a simply defined
emitting region is inadequate to provide all the observed emissions.
Thus we are tempted to invoke another (spatial) region or a second
spectral component.  In almost all cases, this is done with the tacit
assumption that the second component is (or can be) detected in only
one channel: i.e. either synchrotron or IC.  We need to realize that
when we introduce a two zone model, it precludes further analyses
unless there is some hope of observing each zone in both channels.
Some examples of current two zone models are the spine/sheath jet
model \citep[e.g.][]{celo01}; the idea that jets contain regions of
high and low magnetic field strengths, with relativistic electrons
moving between these regions; and the introduction of a
second spectral component to explain hard X-ray \citep{harr99} or
optical \citep{jest05} spectra.}
lead to a total disruption of the jet which maintains its identity
downstream, be it relativistic or not.  What we call knots in quasar
jets, may have little in common with FRI knots given their relative
physical sizes (fig.~\ref{fig:size}).

Insofar as the X-ray emission mechanism 
is concerned, the initial X-ray detection of the Cygnus A hotspots
\citep{harr94} was accompanied with a demonstration that synchrotron
self Compton (SSC) emission provided a consistent explanation if the
average magnetic field strength was close to the equipartition value
under the assumption that the relativistic particle energy density was
dominated by electrons, not protons.  Essentially all the emission
models for jet knots, on the other hand, have shown that SSC emission
is completely inadequate to explain X-ray emission unless the magnetic
field is orders of magnitude less than the equipartition value.

As the number of hotspot detections increased from CXO observations,
many were found to be consistent with SSC predictions but a
significant number appeared to have a larger X-ray intensity than
predicted.  This excess could be attributed to a field strength well
below equipartition, IC emission from the decelerating jet 'seeing'
Doppler boosted hotspot emission \citep{geor03}, or an additional
synchrotron component \citep{hard04b}.  The last named authors show
that the strength of the excess correlates with hotspot luminosity in
the sense that the strongest hotspots are consistent with SSC emission
whereas the weaker radio hotspots required the extra synchrotron
component.

For many distant and/or faint jets, it is often difficult to be
certain that a feature is a knot, a hotspot, or even a lobe.  In some
extreme cases, the true nature of even bright hotspots is ambiguous.
An example is the double hotspot system in 3C351 shown in
fig.~\ref{fig:351}.  Displaced from the north radio lobe is a double
hotspot to the NE of the core.  These are bright at radio and X-ray
bands.  The southern radio lobe has only a weak hotspot with at most
4\% of the radio intensity of the NE hotspots at 1.4 GHz.  Thus the
double hotspot has the hallmarks of relativistic beaming in spite of
the commonly held view that hotspot radiation is not beamed \citep[see
however][for a discussion of beamed emission from hotspots]{denn97}.
Given the fact that these bright features are not located at the outer
edge of the lobe, perhaps they are knots in a jet very close to our
line of sight.

\begin{figure}*
\caption{A radio image of the quasar 3C351 at 1.4 GHz from the VLA.
  X-ray contours are superposed from CXO data.  Contour levels
  increase by factors of 2, from 2 to 32 in arbitrary brightness
  units.   Note the bright NE
  hotspot pair and the very weak S hotspot.}
\label{fig:351}
\end{figure}

\subsection{Entrainment and Collimation}\label{sec:1entrain}

Long-standing problems for low-loss jets include the suppression of
mixing with ambient material and the collimation and stability of jets
\citep{hugh91}.  The process of entrainment of ambient material is
closely related to the process of jet deceleration. Both processes
have been studied observationally \citep[e.g.][]{lain03} and
numerically \citep[e.g.][]{ross04}.  Possible mechanisms causing
entrainment include velocity shear and Kelvin-Helmholtz instabilities
\citep{bodo03}. \citet{lain03} have studied FRI radio galaxies
assuming that the two sides of the jets are intrinsically identical
and that the observed differences in radio brightness and polarization
are caused by the viewing angle and relativistic beaming effects.
They find the velocity of the jet plasma decreases moving away from
the jet axis and that this velocity shear decelerates the jet
substantially.  These arguments purport to demonstrate that there is a
clear distinction between FRI and FRII radio galaxies insofar as their
jet properties are concerned.  Since the powerful jets of FRII radio
galaxies and quasars are able to escape the high ambient density of
their host galaxies and maintain their collimation out to the
prominent hotspots, it is inferred that they suffer less entrainment
and deceleration than FRI jets \citep{bick95}.

Collimation of jets has to be addressed both on sub-parsec scale
during the process of launching the jet and on kpc-scales to explain
the remarkable stability of jets.  \citet{tsin02} for example consider
the former problem and demonstrate collimation for a relativistic
component by a second, non-relativistic less-collimated outflow
(wind). Other collimation mechanisms include confinement by magnetic
fields \citep{saut02}, ram pressure of the ambient medium
\citep{komi94}, and by radiation \citep{fuku01}.

\subsection{Particle Acceleration and Emission Mechanisms}\label{sec:1radi}

The Chandra X-ray observatory increased the number of jets with X-ray
emission from a handful to $\approx$50 sources. Of these, 60\%
are classified as high-luminosity sources (quasars and FRII radio
galaxies) and the remaining are low-luminosity sources (a mix of
FRI's, BL Lac's and a Seyfert galaxy).  The observations indicate that
the radio to X-ray emission from low-luminosity FRI sources can be
explained by synchrotron models, while that from the high-luminosity
FRII sources requires multi-zone synchrotron models, synchrotron and
IC models, or more exotic variants.
\subsubsection*{Synchrotron models for FRI galaxies}
For low-luminosity (FRI) radio sources, there is strong support for
the synchrotron process as the dominant emission mechanism for the
X-rays, optical, and of course radio emissions.  Among the arguments
supporting this view are the intensity variability found for knots in
the M87 jet \citep{harr06a}; the fact that in most cases the X-ray
spectral index, $\alpha_x$ is $>$~1 and significantly larger than the
radio index, $\alpha_r$; and the relative
\marginnote{{\bf Energy Losses \& Halflives} Relativistic electrons
loose energy via several processes.  For both synchrotron and inverse
Compton radiation, the rate of energy loss is $\propto$ E$^2$ (E is
the electron's energy).  For these loss channels, the time it takes to loose
half the energy ('half-life') is $\propto$ E$^{-1}$.}
morphologies in radio, optical, and X-rays.  For the sorts of magnetic
field strengths generally ascribed to jet knots (10 to 1000 $\mu$G),
synchrotron X-ray emission requires the presence of electrons of energies in
the range 10$^7~<~\gamma~<~10^8$.  As the highest energy electrons
cool in equipartition magnetic fields on time scales of years, the
observations of single power-law spectral energy distributions
extending all the way from the radio to the X-ray regime pose the
problem of why there is no sign of radiative cooling.  A possible
solution may be that electrons escape the high-magnetic field
emission region before they cool.

The radio to X-ray observations require the presence of one or more
populations of high energy electrons (or protons if proton synchrotron
emission is viable, \citep{ahar02}).  A common assumption is that the
particles are accelerated at strong magnetohydrodynamic (MHD) shocks
by the Fermi I mechanism \citep{bell78,blan78}\citep[see also the
review by][]{kirk99}.  However, there are several uncertainties.
First we cannot be sure that the Fermi process is relevant since if
the jet is strongly magnetized with a tangled field geometry, shock
acceleration is not as effective as for strong shocks which can exist
when the field does not dominate.  Next, the uncertainty of the bulk
Lorentz factor of the jet medium means that we can't be sure that
$\Gamma$ is large enough to allow the possibility of relativistic
shocks.  Finally, even if the bulk velocity of the jet is
relativistic, it is still possible to have non-relativistic shocks in
the jet frame.  For mildly relativistic shocks, Fermi I shock
acceleration is more complicated than is the case for the
non-relativistic regime, and it is not yet well understood
\citep{kirk99}).


\paragraph{Distributed Acceleration}\label{sec:1distrib}

For FRI jets such as that in M87 (fig.~\ref{fig:m87}),
X-ray emitting electrons with $\gamma\approx~10^7$ will cool on time
scales of a few years, and optical and UV emitting electrons will cool
on time scales of a few decades.  Thus, interpreting the X-ray and
optical emission from these sources as synchrotron emission implies
that the emitting regions cannot be much larger than the electron
acceleration regions. For bright knots which have traditionally been
associated with strong shocks in the jet flow, these ``life-time
constraints'' can easily be accommodated.  However, Chandra detected
several jets with quasi-continuous emission along the jet
\citep[e.g. Cen A,][]{kata06} suggesting that electron acceleration may be
spatially distributed rather than being restricted to a few bright
knots.  \citet{wang02} finds that plasma turbulent waves can be a
mechanism for efficient particle acceleration, producing high energy
electrons in the context of blazar jets.  \citet{nish05,staw02}
propose turbulent acceleration in a jet's 'boundary' or 'shear' layer
surrounding the jet spine.  \citet{staw02} also argue that the resulting
electron energy distribution should show an excess near the high
energy cutoff, thereby producing a harder X-ray spectrum than would be
expected based on the extrapolation of the radio and optical data.

Other explanations for the quasi-continuous emission include low-level
IC/CMB emission of low-energy electrons (see the discussion in the
next paragraph) and synchrotron emission from electrons accelerated by
magnetic reconnection.  However, if the knot emission is produced by
synchrotron emission from shock-accelerated electrons and the
continuous emission has another origin, one might expect that the two
jet regions would show markedly different spectral energy
distributions. Measurement of X-ray spectral indices of the continuous
emission is usually difficult, because of the fewer photons available
for analysis. In the case of M87, \citet{perl03} find no change of
$\alpha_x$ between the knots and the quasi-continuous emission within
a statistical accuracy of $\pm$0.15 in the X-ray spectral index.

\begin{figure}
\caption{M87 CXO image with 8 GHz contours.  The X-ray image has an
  effective exposure of about 115 ks, consisting of 22 observations
  taken between 2000 and 2004. It has been smoothed with a
  Gaussian of FWHM=0.25$^{\prime\prime}$ and the energy band is 0.2-6
  keV.  The color mapping is logarithmic and ranges from 0.02 (faint
  green) to a peak of 5.5 ev s$^{-1}$ (0.049$^{\prime\prime}$
  pixel)$^{-1}$.  The radio data are from the VLA with a beam of
  FWHM=0.2$^{\prime\prime}$.  Contour levels increase by factors of 2
  and start at 1 mJy/beam.}
\label{fig:m87}
\end{figure}

\paragraph{Departures from power law spectra}\label{sec:1bowtie}

\begin{figure}
 \centerline{\psfig{figure=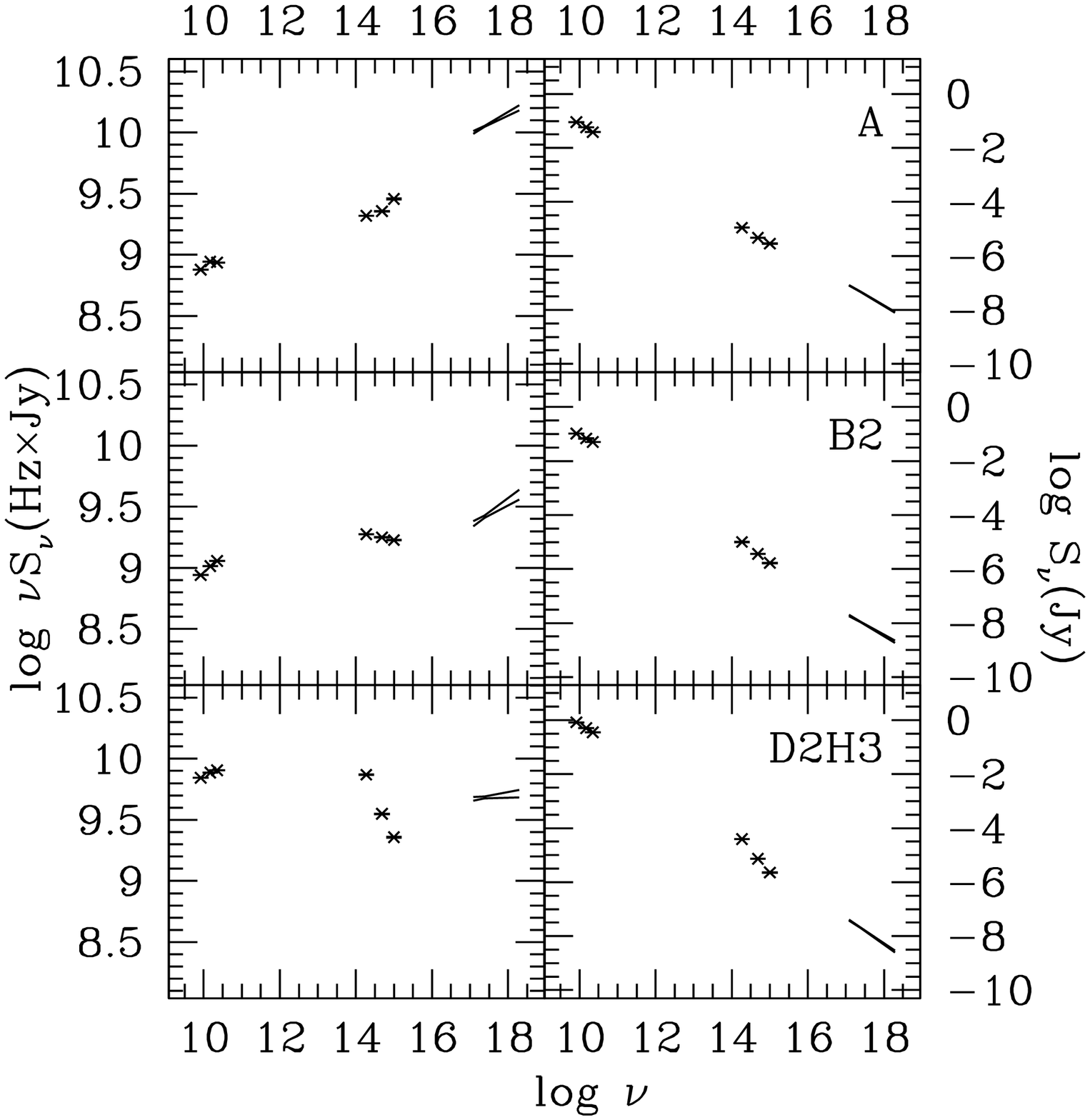,height=30pc}}
\caption{Examples of spectral energy distributions: 3 knots in the
  3C273 jet.  The left panels
  plot log($\nu\times$S$_{\nu}$) as ordinate and the right panels are
  versions with logS$_{\nu}$ vs. log$\nu$.  From top to bottom knots
  A, B2, and D2/H3 are shown.  The X-ray data are presented as
  ``bowties'' which delineate the range of acceptable power-law
  slopes.  Note how these data preclude a fit consisting of a single
  zone synchrotron spectrum which would be a curve concave downward.
  This figure was kindly provided by S. Jester.  Details can be found
  in \citet{jest06}.}
\label{fig:2sed}
\end{figure}

One of the primary reasons that IC/CMB models are preferred over
synchrotron models for most FRII radio galaxies and quasars is the
so-called `bow-tie problem'.  Conventional synchrotron spectral energy
distributions call for a concave downward spectral shape, allowing for
spectral breaks to steeper spectra at higher frequencies and eventual
\marginnote{{\bf SED:}
Spectral Energy Distribution: To describe the continuum spectrum of
a feature, log($\nu\times$flux density) is plotted against
log$\nu$.  We use the term also for log(flux density) vs. log$\nu$.}
high frequency cutoffs.  Thus we expect $\alpha_x~\geq~\alpha_{ox}$
($\alpha_{ox}$ is the spectral index between optical/UV and X-ray).
When this is not the case, the 'bow-tie' showing the X-ray flux
density and allowed range of $\alpha_x$ does not permit a smooth fit
of a concave downward curve and instead requires a flattening of the
X-ray spectrum.  Examples are provided in fig.~\ref{fig:2sed} which
shows the SED's for 3 knots in the 3C273 jet.  The bow-tie problem is
more common for FRII radio galaxies and quasars but is also found for
some of the FRI radio galaxies.

In addition to the hypothesis of a second spectral component, there have
been two suggestions for accommodating this behavior with synchrotron
emission.  The first is that mentioned above: boundary layer
acceleration \citep{staw02} producing a flatter spectrum for the high
energy part of the electron spectrum.  The other suggestion is
restricted to the case where IC dominates the E$^2$
losses. \citet{derm02} argue that for the highest energy electrons, IC
losses are reduced by the lower Klein-Nishina cross section so that
the top end of the electron distribution experiences a reduced loss
rate and thus an excess above the expected distribution builds up at
high energies producing a hard synchrotron spectrum at X-ray
frequencies.  Although this is a clever method of solving the bow-tie
problem, in order to work, the photon energy density in the jet frame,
u$^{\prime}$($\nu$) must be larger than the magnetic field energy
density, u(B).  To realize this, a $\Gamma^2$ boosting of the CMB is
required, and the practical result is that by invoking the necessary
$\Gamma$, you will already produce the observed X-ray emission by the
IC/CMB process.

\subsubsection*{Emission models for FRII galaxies and quasars}
The most pressing problem for X-ray emission from relativistic jets is
the emission mechanism for the powerful jets from FRII radio galaxies
and quasars.  As mentioned above, the radio to X-ray spectral energy
distributions of most of these sources cannot be described by a
one-component synchrotron model.  Such models predict a spectral
energy distribution which softens at high energies. In terms of
spectral indices, we expect $\alpha_x\geq\alpha_{ox}$), whereas the
Chandra observations showed that $\alpha_x<\alpha_{ox}$ for many
quasar jets.  The most popular explanation is the IC model put forth
by \citet{celo01} and \citet{tave00}.  The observed large ratios of
X-ray to radio luminosities are explained by postulating very fast
jets with high bulk Lorentz factors $\Gamma$. Relativistic boosting
increases the energy density of the CMB in the jet frame:
\begin{equation}
u'(\rm CMB)\,= 4 \times 10^{-13}~(1+z)^4~\Gamma^2~\rm erg~cm^{-3}.
\end{equation}
In this way, a single population of electrons is able to
produce the radio and optical synchrotron emission in a magnetic field
close to equipartition (B$_{eq}$ generally less than 100 $\mu$G), and
the IC X-ray emission by scattering off the relativistically boosted
CMB.

\begin{figure}
\centerline{\psfig{figure=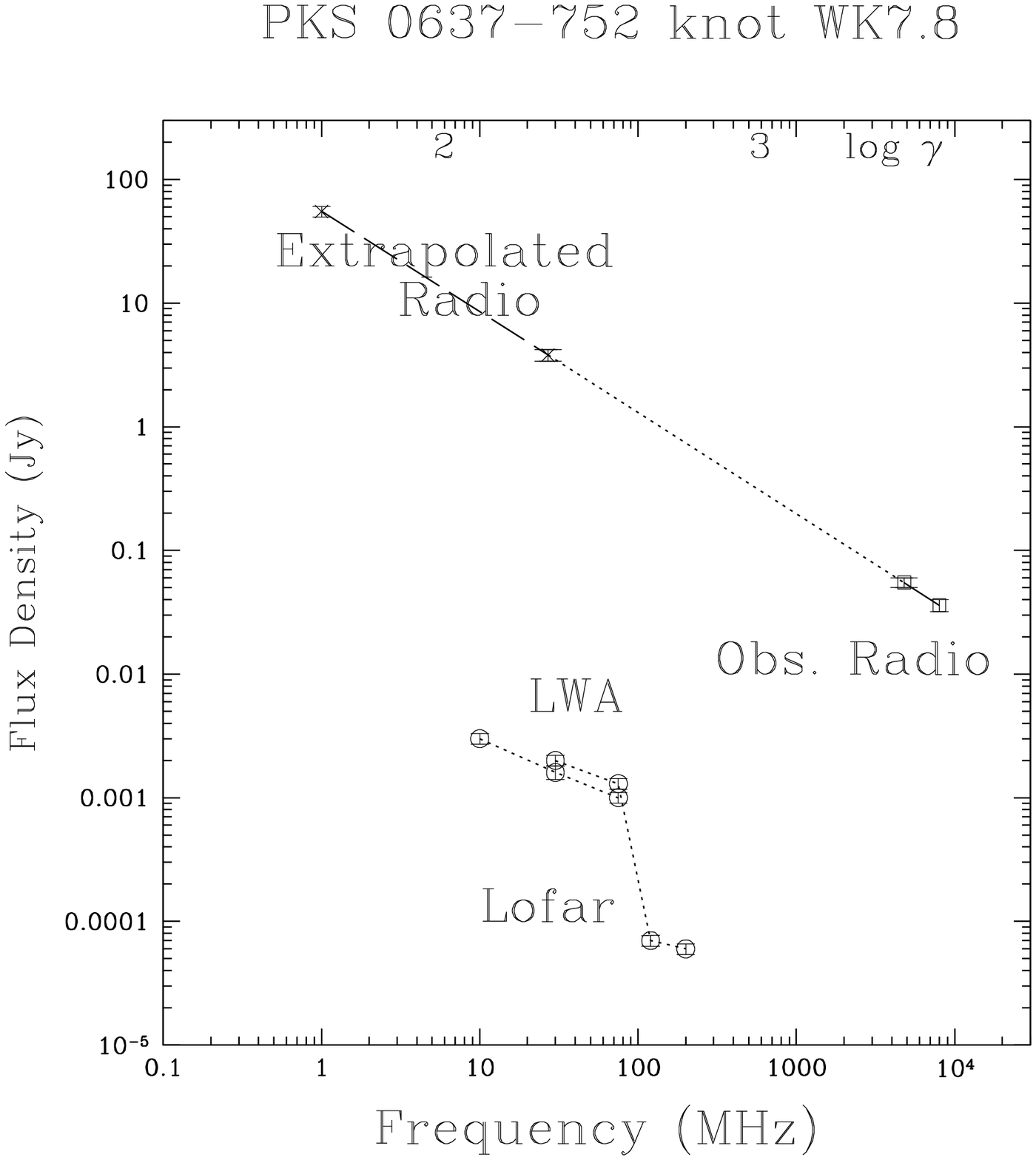,height=25pc}}
\caption{Segments of the synchrotron spectrum of knot WK7.8 in the jet
  of PKS0637-752.  The observed radio flux densities are plotted toward
  the right edge so the short solid line allows us to determine a
  small section of the electron spectrum if we trust the equipartition
  field strength estimate; the extrapolation of this spectrum to lower
  frequencies is shown by the dotted line.  The segment coming from
  the same electrons responsible for the IC/CMB X-ray emission is the
  long dash line to the upper left.  Also shown near the bottom are
  sensitivity limits for low frequency radio telescopes being designed
  and under construction: LWA is the Long Wavelength Array planned for a
  site in the southwest of the US, and LOFAR is being built in the
  Netherlands. At the top are a few values of log $\gamma$ which
  indicates the energy of the electrons corresponding to the emission
  spectrum.}
\label{fig:extrap}
\end{figure}

While we evaluate the various difficulties confronting the IC/CMB
model in the final section of this review (\ref{sec:4ic}), we would
like to emphasize here that the IC/CMB model requires two
key-ingredients for which there is at present no independent
observational verification: enough low-energy electrons and highly
relativistic plasma motion on kpc-scales.  Analysis of the spectral
energy distributions of several FRII sources shows that electrons with
Lorentz factors $\gamma'~\approx$~100 produce the observed X-rays:
\citep[e.g.][eq. B4]{harr02a}:

\begin{equation}
\gamma' \,=\, \sqrt{\frac{6.25\times 10^{-12}\, \nu_{\rm
    ic}(obs)}{(1+\mu'_{\rm j}) \delta \Gamma}}.  
\label{eq:gammaic}
\end{equation}

\noindent
where the prime is used to denote quantities in the jet-frame,
$\mu'_{\rm j}$ equals cos($\theta'$) and $\theta'$ is the angle
between the jet direction and the line of sight. The uncertainties of
extrapolating the electron spectrum to low energies is illustrated in
fig.~\ref{fig:extrap} which shows the spectrum of the knot in the jet
of PKS0637-752. The low-energy electrons responsible for the X-ray
emission produce synchrotron emission in the 1-30 MHz range, well
below frequencies available from the Earth with reasonable angular
resolutions. For this example we have used $\Gamma$=10, which is the
value required for the IC/CMB model \citep{tave00,celo01}.  The actual
electron spectrum could flatten significantly for $\gamma\leq$3000 or
even suffer a low energy cutoff.  In that case there would be fewer
electrons than calculated, and the required value of $\Gamma$ would
have to be increased to compensate.  It is even conceivable that the
electron spectrum could steepen at low energies and the required
$\Gamma$ would be much smaller than estimated.  Our ignorance of the
low end of the electron spectrum is very general; only in sources with
very high values of magnetic field strength do ground based radio data
begin to give us the required information.

On the positive side, the IC/CMB model avoids the
'far-from-equipartition' requirement of models which explain the high
X-ray fluxes as synchrotron self-Compton emission from electrons
up-scattering long-wavelength synchrotron photons into the X-ray band
\cite[e.g.][]{schw00}.  Furthermore, it does not require the ad-hoc
introduction of additional particle components, required by multi-zone
synchrotron models \cite[e.g.][]{harr99}.


\section{Physical Comparisons of Resolved X-ray Jets}\label{sec:phys}

We are used to looking at images of jets that fit nicely on the page,
be they Galactic micro quasars, jets from relatively local FRI radio
galaxies, or jets from quasars with substantial redshifts.  We are
struck by a number of similarities and are tempted to consider them
all to share fundamental properties.  In an effort to sharpen our
perspective, we have devoted this part of the review to presenting the
observed and deduced parameters for the X-ray jets known to us.  Most
of these data exist in the literature, but we have adjusted published
values, where necessary, to conform to the currently standard
cosmology: H$_0$=71~km~s$^{-1}$; $\Omega_m$=0.3; and
$\Omega_{\Lambda}$=0.7.  Gathering these data will allow us to compare
physical sizes and apparent luminosities.

In fig.~\ref{fig:size} we show the relative sizes of 3 jets: M87,
3C273, and PKS1127-145.  Although the indicated sizes (1.6, 56, and 238
kpc, respectively) are projected sizes, it can be seen that the entire
jet of M87 would easily fit within a single knot of the 3C273 jet.
Note also that a single 0.049$^{\prime\prime}$ pixel in the top panel
corresponds to a few pc, the scale of VLBI jets and also comparable to
the total size of jets from microquasars in our galaxy.  Given the
vast range of scales, can we really expect similar physical processes
to operate all the way from pc to Mpc scales?

\begin{figure}
\caption{A comparison of 3 jets.  The X-ray jets, M87 (top), 3C273
  (middle), and PKS1127-145 (bottom) have been rotated for ease of
  comparison.  All maps have had pixel randomization removed and have
  been smoothed with a Gaussian of FWHM=0.25$^{\prime\prime}$.  The
  absolute brightness mapping is logarithmic and the same for all
  three and ranges from 0.01 (pink) to 6.7 (the black peak of 3C273).
  The intensity units are total electron volts per sec per pixel and
  the pixel size is 0.0492$^{\prime\prime}$.  The overall projected
  size of each jet (from top to bottom) is 1.6, 56, and 238 kpc
  (21$^{\prime\prime}$, 21$^{\prime\prime}$, and 29$^{\prime\prime}$,
  respectively).  The small cyan line overlaid on the core of 3C273
  shows the length of the M87 jet if it were moved to the distance of
  3C273, and the cyan rectangle on the bottom panel represents the
  total length of the 3C273 jet if it were at the distance of
  PKS1127-145.  The X-ray jet of PKS1127-145 is too faint to be
  visible on the common intensity scale adopted, so radio contours are
  overlaid in order to show the full extent of the jet.  X-ray
  emission is detected out to the last radio feature
  (fig.~\ref{fig:1127}).}
\label{fig:size}
\end{figure}

\subsection{Gross jet properties}\label{sec:2jets}

In Table~\ref{tab:fr1jets} we list parameters for X-ray jets of low
radio power sources from the XJET
website\footnote{http://hea-www.harvard.edu/XJET/} (2005.6), and
Table~\ref{tab:fr2jets} contains the data for the more powerful
sources classified as quasars or FRII radio galaxies.  We have not
included sources if only the terminal hotspots and/or lobes have been
detected in X-rays.  In almost all cases, the division between the two
tables corresponds to how the original investigators interpret the
X-ray emission process.  All the entries in Table~\ref{tab:fr1jets}
are ascribed to synchrotron emission except for Cen B, and similarly
the jets of Table~\ref{tab:fr2jets} are described on the basis of the
IC/CMB model except for PKS2152-69 for which \citet{ly05} suggested
thermal emission; Pictor A, PKS 1136-135 and 3C 273 for which both
synchrotron and IC/CMB have been suggested; and 1928+738 and 3C403
which have been ascribed to synchrotron emission.

The projected jet length, both in arcsec and kpc should be accurate to about
10\%; it is meant to describe the length of the X-ray jet as detected
by the CXO and not the
total length of the radio jet.  The apparent X-ray luminosity is derived from
the observed flux or flux density, assuming $\alpha_x$=1.  As pointed
out by \citet{list03a}, such luminosities are not directly useful for
correlations since we are dealing with relativistic beaming which
alters the jet frame luminosity, depending on $\Gamma$ and $\theta$.  The
value of $\alpha_x$ given is a published value, either for the whole
jet, or from a brighter knot.  If a reasonable estimate of the angle
of the jet to the line of sight is given in the literature, it is
quoted here.  For a number of the quasars, $\theta$ is estimated from
the IC/CMB calculation, and is thus model dependent; for others, it is
estimated from VLBI studies.  The resulting deprojected
length suffers from similar uncertainties.

\begin{table}
\def~{\hphantom{0}}
\caption{Parameters for Jets of Low Power Radio Galaxies}
\label{tab:fr1jets}
\scriptsize

\begin{tabular}{@{}lllcrllccl@{}}
\toprule
Host name  &  z  &  Scale  & Length  &  Length  &  log~L$_x$ &
$\alpha_x$  &  $\theta$  &  Deproj.  &   Reference\\
 & & (kpc/'') & (arcsec) & (kpc) & (ergs$^{-1}$) & & (deg) & (kpc)  & \\  
\colrule

 3C15*      & 0.0730  & 1.4  & 4  & 5.6        & 41.03  & 0.7$\pm$0.4  & ..  & ..  & 1 \\
 NGC 315    & 0.0165  & 0.33  & 13  & 4.3   & 40.54  & 1.5$\pm$0.7  & ..  & ..  & 2 \\
 3C31*      & 0.0167  & 0.34  & 8  & 2.7       & 40.56  & 1.1$\pm$0.2  & 52  & 3.4  & 3   \\
 B2 0206+35 & 0.0369  & 0.72  & 2  & 1.4 & 41.12  & ..  & ..  & ..  &  4  \\
 3C 66B*    & 0.0215  & 0.43  & 7  & 3.0     & 41.03  & 1.3$\pm$0.1  & ..  & ..  & 5 \\
 3C 120*    & 0.0330  & 0.65  & 80  & 52     & 41.95  & ..  & ..  & ..  &  6 \\
 3C 129     & 0.0208  & 0.42  & 2.5  & 1.0   & 39.64  & ..  & ..  & ..  & 7 \\
 PKS 0521-365* & 0.055  & 1.06  & 2  & 2.1    & 41.90  & 1.2$\pm$0.3  & ..  & ..  & 8 \\
 B2 0755+37*& 0.0428 & 0.83  & 4  & 3.3 & 41.52  & ..  & ..  & ..  &  4  \\
 3C270      & 0.0074 & 0.15  & 35  & 5.2    & 39.13  & ..  & ..  & ..  &  9, 10 \\
 M84        & ...    & 0.082  & 3.9  & 0.3        & 38.71  & 0.8$\pm$0.3  & 50  & 0.4   & 11 \\
 M87*       & 0.0043 & 0.077  & 20  & 1.5     & 41.32  & $>$1  & 20  & 4.5    & 12, 13 \\
 Cen A      & ...    & 0.017  & 120  & 2.0      & 39.39  & 0.4 to 2.2  & 15  & 7.7  &  14 \\
 Cen B      & 0.013  & 0.26  & 8  & 2.1      & 40.13  & ..  & ...  & ...  & 15 \\
 3C296      & 0.0237 & 0.47 & 10 & 4.7        & 40.09 & 1.0$\pm$0.4 & .. & .. &  16 \\
 NGC6251*   & 0.0249 & 0.49  & 410  & 200 & ..  & 1.30$\pm$0.14 & ..  & ..  & 17 \\
 3C 346*    & 0.161  & 2.7  & 2  & 5.4      & 41.96  & 1.0$\pm$0.3  & 20  & 16  & 18 \\
 3C 371*    & 0.051  & 0.98  & 4  & 3.9      & 41.87  & 0.7+0.4,-0.2  & 18  & 12.6  & 19 \\
 3C 465     & 0.0293 & 0.58  & 7.5  & 4.4  & 40.30  & $\approx$~1.4  & ...  & ...  & 16\\

\botrule
\end{tabular}
Notes\\
The scale is given in units of kpc per arcsec.\\

All sources are classified as FRI radio galaxies except for 3C120, a
Seyfert I galaxy, and the two BL Lac objects, PKS0521-365 and 3C371.\\

A ``*'' after the source name indicates that an optical detection has
been reported, see http://home.fnal.gov/$\sim$jester/optjets/.\\
 
References\\ 1 \citet{kata03}; 2 \citet{worr03}; 3 \citet{hard02}; 4
\citet{worr01}; 5 \citet{hard01}; 6 \citet{harr04}; 7 \citet{harr02c};
8 \citet{birk02}; 9 \citet{zeza05}; 10 \citet{chia03}; 11
\citet{harr02b}; 12 \citet{wils02}; 13 \citet{mars02}; 14
\citet{hard03}; 15 \citet{mars05}; 16 \citet{hard05a}; 17
\citet{evan05}; 18 \citet{worr05}; 19 \citet{pesc01}.  

\normalsize
\end{table}


\begin{table}
\def~{\hphantom{0}}
\caption{Parameters for Jets of High Power Radio Galaxies \& Quasars}
\label{tab:fr2jets}
\scriptsize
\begin{tabular}{@{}lllcrllccrl@{}}
\toprule

Host name  &  z  &  Scale  & Length  &  Length  &  log L$_x$ &
$\alpha_x$  &  $\theta$  &  Deproj.  &  $\delta$ &  Reference\\
 & & (kpc/as) & (arcsec) & (kpc) &
(ergs$^{-1}$) & & (deg) & (kpc) &   & \\  

\colrule
3C9           &  2.012  & 8.5  & 6.4  & 54           & 44.34  & ..  & ..  & .. & .. & 1 \\
 PKS 0208-512 &  0.999  & 8.04  & 5  & 40     & 44.47  & ..  & 8  & 262 & 7  & 2,3 \\
 PKS 0413-21  &  0.808  & 7.54  & 2  & 15     & 43.99  & ..  & 20  &44  & 3 & 2 \\
 Pictor A     & 0.0350  & 0.69  & 114  & 79   & 40.84  & 0.97$\pm$0.07   & $>$23  & $<$201  & $<$3 & 4,5 \\
 PKS 0605-085 &  0.870  & 7.7  & 4  & 31      & 44.58  & 0.4$\pm$0.7  & ..  &.. & ..  & 6  \\
 PKS 0637-752* & 0.651  & 6.9  & 12  & 83      & 44.34  & 0.85$\pm$0.08 & 5.7  & 836  & 10  & 7,8 \\
 3C 179       & 0.846  & 7.7  & 4.4  & 34       & 44.45  & ..  & ..  & ..  & .. & 6\\
 B2 0738+313  & 0.635  & 6.9  & 35  & 241  & 42.93  & 0.5 to 1.4  & 8  & 1730  & 7  & 9 \\
 0827+243     & 0.939  & 7.9  & 6.2  & 49     & 44.14  & 0.4$\pm$0.2  & 2.5  & 1100  & 20  & 10 \\
 3C 207       & 0.68  & 7.1  & 4.6  & 33        & 43.97  & 0.3$\pm$0.3  & 8  & 237  & 7  & 6\\
 3C 212*      & 1.049  & 8.1  & 4  & 32         & 43.52  & ..  & ..  & ..  & ..  &  11  \\
 PKS 0903-57  &  0.695  & 7.1  & 3.5  & 25    & 43.90  & ..  & 20  & 73  & 3  & 2 \\
 PKS 0920-39  &  0.591  & 6.6  & 10  & 66     & 43.70  & ..  & 7  & 322  & 8  & 2,3 \\
 3C 219       & 0.174  & 2.9  & 20  & 58        & (41.68)  & ..  & ..  & .. & ..  & 12 \\
 Q0957+561    & 1.41  & 8.5  & 8  & 68       & 43.69  & 0.9$\pm$0.6  & ..  & ..  & 1.4  & 13 \\
 PKS 1030-357 &  1.455  & 8.5  & 12  & 102    & 44.99  & ..  & 8.6  & 682  & 9  & 2,3  \\
 PKS 1046-40  &  0.620  & 6.8  & 4  & 27      & 43.44  & ..  & 17  & 93  & 3  & 2 \\
 PKS 1127-145 & 1.18  & 8.3  & 30  & 249      & 44.62  & 0.5$\pm$0.2  & 24  & 612  & 4  &  14  \\
 PKS 1136-135*& 0.554  & 6.4  & 6.7  & 43     & 43.92  & 0.4$\pm$0.4  & 6  & 410  & 10  & 6 \\
 4C49.22*     & 0.334  & 4.8  & 5.6  & 27      & 43.62  & ..  & 6  & 270  & 14  & 6 \\
 PKS 1202-262 &  0.789  & 7.5  & 5  & 37      & 44.73  & ..  & 4.9  & 568  & 12  & 2,3  \\
 3C 273*      & 0.1583  & 2.7  & 21  & 57       & 43.58  & 0.6 to 0.9 & 5  & 654  & 5  & 16,17\\
 4C19.44*     & 0.720  & 7.2  & 14.4  & 104    & 44.55  & ..  & 10  & 616  & 14  & 6 \\
 3C 303*      & 0.141  & 2.5  & 9  & 22         & 41.56  & ..  & ..  & ..  & ..  & 17  \\
 GB 1508+5714 & 4.3  & 6.9  & 2.2  & 15        & 44.96  & 0.9$\pm$0.4  & 15  & 58  & 4  & 18,21 \\
 PKS 1510-089 &  0.361  & 5.0  & 5.2  & 26    & 43.85  & 0.5$\pm$0.4  & ..  & ..  & ..  & 6 \\
 3C 345*      &  0.594  & 6.6  & 2.7  & 18      & 43.65  & 0.7$\pm$0.9  & 7  & 138  & 7  & 6 \\
 1642+690     &  0.751  & 7.3  & 2.7  & 20    & 43.67  & ..  & ..  & ..  & ..  & 6 \\
 3C 380*      &  0.692  & 7.1  & 1.8  & 13      & 44.68  & ..  & 13  & 57  & ..  & 2 \\
 1928+738*    &  0.302  & 4.4  & 2.6  & 11    & 43.21  & 0.7$\pm$0.7  & 6  & 105  & 10  & 6 \\
 3C403*       & 0.059 & 1.13 & 45 & 51            & 41.48 & 0.7$\pm$0.4 & .. & .. & .. & 19  \\
 PKS 2101-490 &  (1.04)  & 8.1  & 6  & 49        & 44.17  & ..  & 25  & 116  & ..  &  2 \\
 PKS 2152-69  &  0.0283  & 0.56  & 10  & 5.6  & 40.66  & 1.6$\pm$0.4  & ..  & ..  & ..  & 20 \\
 3C 454.3*    &  0.859  & 7.7  & 5.2  & 40    & 44.62  & ..  & 18  & 129  & ..  &  2 \\
\botrule
\end{tabular}
Notes\\
The scale is given in units of kpc per arcsec.\\

All sources are classified as quasars except for the 4 FRII radio galaxies:
Pictor A, 3C219, 3C403, and PKS2152-69.\\

A ``*'' after the source name indicates that an optical detection has
been reported see http://home.fnal.gov/$\sim$jester/optjets/.\\

The redshift for PKS 2101-490 is uncertain (described as 'tentative' by \citet{mars05}).\\

PKS2152-69 is odd, having a bright knot close to the core, and a
disparity between the radio, optical, and X-ray distributions.
\citet{ly05} argue for a thermal interpretation of the X-ray emission.\\

References\\
1, \citet{fabi03}; 2, \citet{mars05}; 3, \citet{schw06}; 4,
\citet{wils01}; 5, \citet{hard05b}; 6, \citet{samb04}; 7,
\citet{char00}; 8, \citet{schw00}; 9, \citet{siem03a}; 10,
\citet{jors04a};11, \citet{aldc03}; 12, \citet{coma03}; 13,
\citet{char02}; 14, \citet{siem02}; 15, \citet{mars01}; 16,
\citet{samb01}; 17, \citet{kata03}; 18, \citet{cheu04}; 19, \citet{ kraf05}; 20, \citet{ly05}; 21, \citet{siem03b}.   
\normalsize
\end{table}

\normalsize

\subsection{Evaluation}\label{sec:2eval}

\begin{figure}
\centerline{\psfig{figure=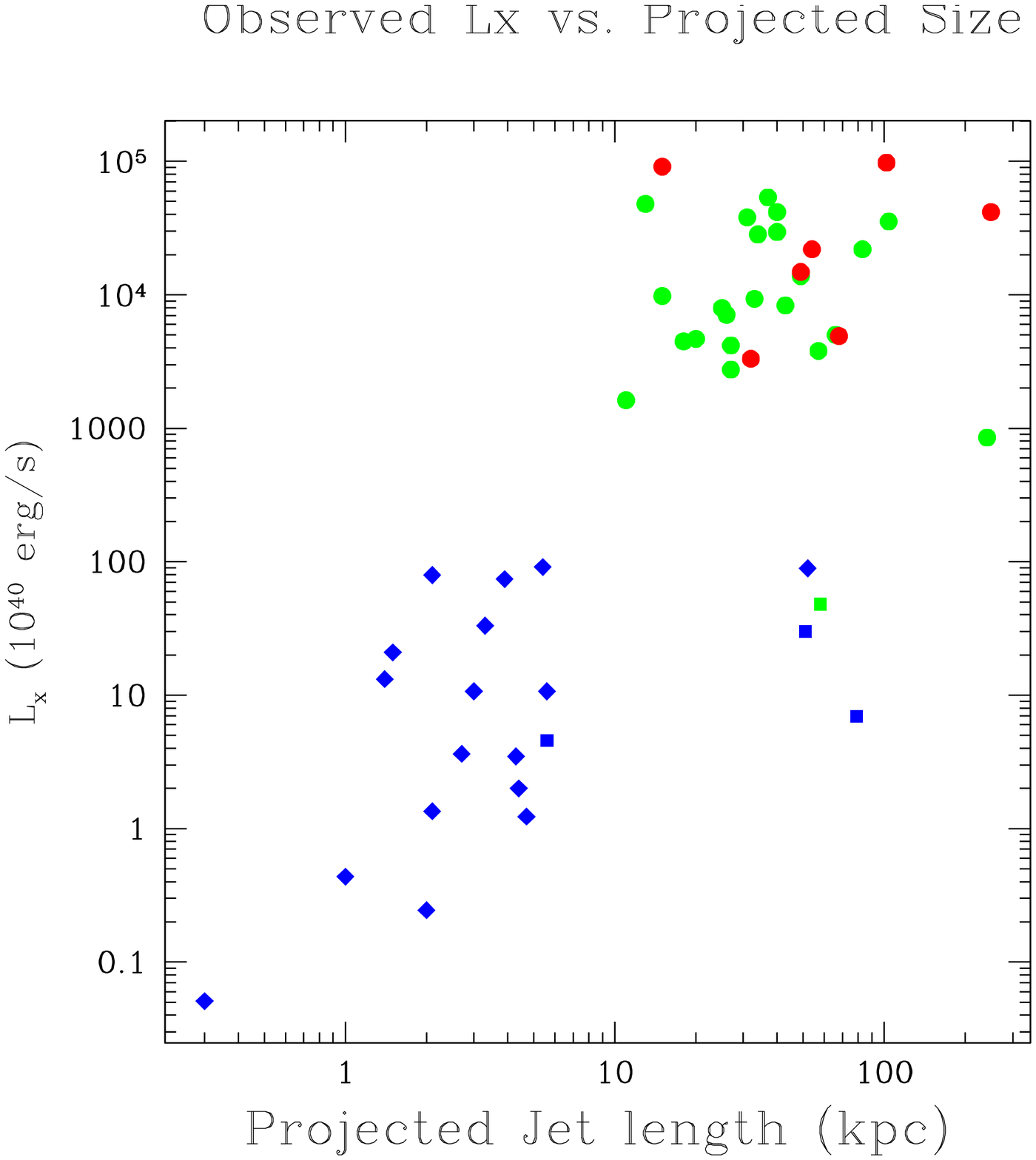,height=30pc}}
\caption{The observed X-ray luminosity plotted against the projected
  length of the jet.  Quasars are plotted with filled circles; FRII
  radio galaxies with squares; and FRI radio galaxies (including
  Seyferts and BL Lac objects) with diamonds.  The colors are
  allocated according to: red, z$>$1; green, 1$>$z$>$0.1; and blue,
  0.1$>$z.  The FRI (diamond) close to the 3 FRII's is 3C120 which has
  a weak detection of a knot 80$^{\prime\prime}$ from the nucleus.
  The FRII (square) jet in the midst of the main clump of FRI's is
  PKS2152-69.  \citet{ly05} present evidence that the X-ray emission of
  the jet related feature
  is thermal in origin.}
\label{fig:obs_phys}
\end{figure}

In fig.~\ref{fig:obs_phys} we show a plot of the observed parameters,
jet length (projected) and observed (i.e. assuming isotropic emission)
X-ray luminosity, L$_x$.  This plot conforms to the common perception
that quasars have powerful jets and are generally longer than those of
FRI galaxies.  Perhaps the only surprise is the gap with no jets lying
between 10$^{42}$ and 10$^{43}$~ergs~s$^{-1}$.  The lower right is
sparsely populated partly because in a large fraction of FRI jets,
only the inner segment is detected in X-rays.  The upper left is empty
because short jets at typical quasar redshifts will be difficult to
resolve from the nuclear emission with arcsec resolutions.  A
separation of $\approx$2$''$ is required to detect a jet close to a
bright quasar, and at a typical redshift of 0.5, this already
corresponds to 10kpc.

The FRII radio galaxies have projected sizes comparable to those of
the quasars, but are of lower apparent luminosity.  The weakest jet
(lower left corner) is M84 for which X-ray emission has been detected
in the very inner part of the radio jet.  3C129 is quite similar, and
joins Cen A, both points lying to the lower left of the main clump of
FRI's.

This sort of plot is useful for comparative purposes, but not for
interpretation because L$_x$ is only an apparent luminosity and not
the true luminosity in the jet frame and also because the length is a
lower limit because of projection.

\begin{figure}
\centerline{\psfig{figure=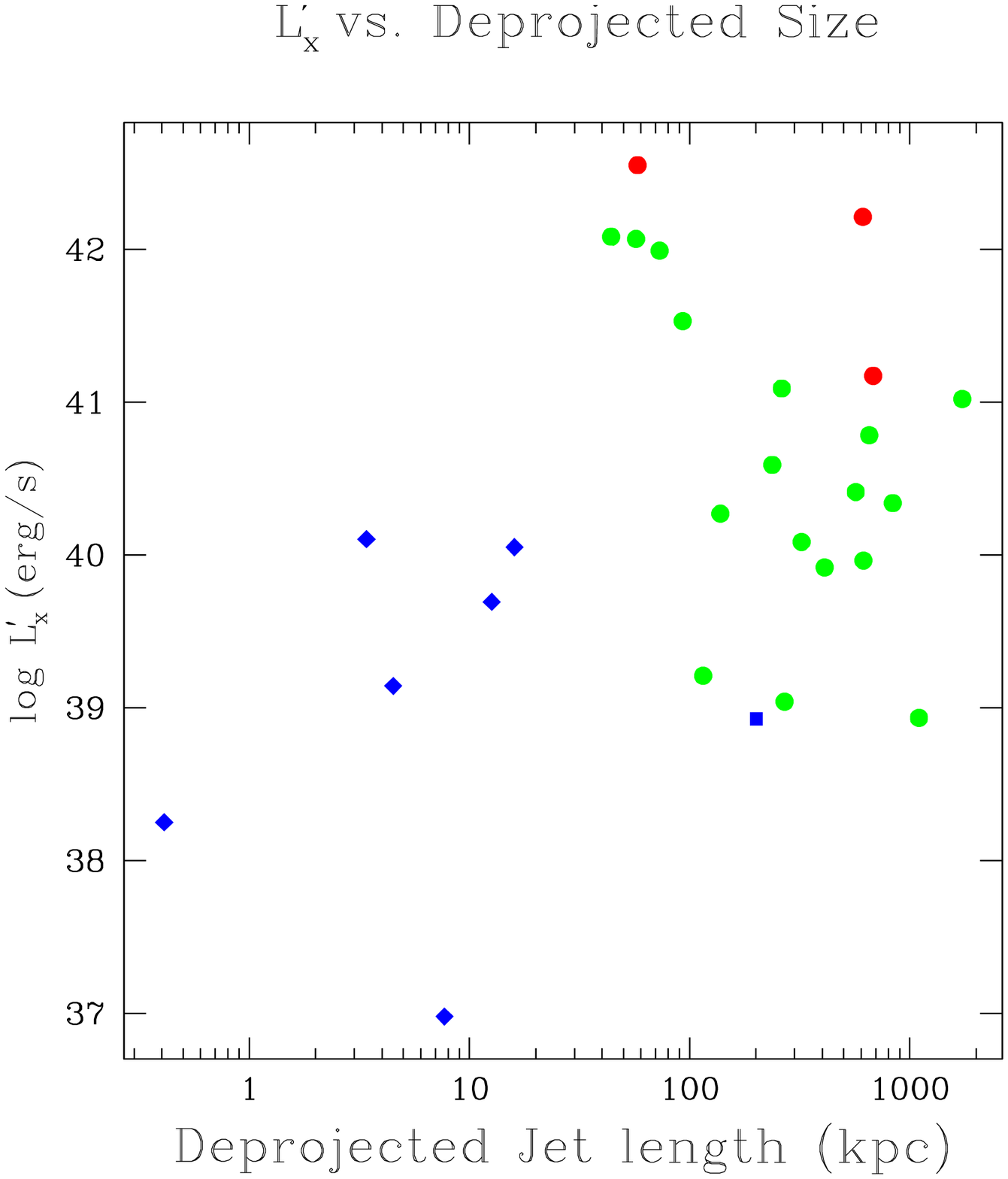,height=30pc}}
\caption{The best guess jet-frame luminosity corresponding to the
  observed X-ray luminosity, (L$_x^{\prime}$=L$_x$(obs)/$\delta^4$),
  plotted against the deprojected length of the jet.  The symbols are
  used the same way as in fig.~\ref{fig:obs_phys}.}
\label{fig:lxp_phys}
\end{figure}

For a subset of the sources plotted in fig.~\ref{fig:obs_phys}, some
reasonable estimate for the angle between the line of sight and the
jet has been published.  Since most of these jets are sensibly
straight on kpc scales (3C120 being a notable exception), we can
obtain a deprojected length.  For most of these, an estimate of the
beaming factor is also available.  For the majority of the quasars,
the value of $\delta$ given in Table~\ref{tab:fr2jets} is model
dependent since it is the beaming factor required for the IC/CMB
model.  For the FRI and FRII radio galaxies, the $\delta$ values are
derived from various lines of arguments based on geometry of the
lobes, VLBI superluminal motions, and other more or less reliable
methods.  Assumed $\delta$'s for the FRI jets are: 1.3, 3C31 and M84;
3.5, M87 and 3C371; 4, Cen A; and 3, 3C346.  However, all beaming
factors are suspect and the corresponding uncertainty will most likely
introduce scatter in plots such as fig.~\ref{fig:lxp_phys} which plots
L$_x^{\prime}$=L$_x(obs)/\delta^4$ against length(obs)/sin$\theta$.

The main purpose of fig.~\ref{fig:lxp_phys} is to demonstrate that
with the 'current community interpretation' (i.e. FRI jets come from
synchrotron emission whereas quasar jets are dominated by IC/CMB
emission), FRI jets and quasar jets are more clearly separated on the
basis of size rather than luminosity.  Parameters for the smallest
quasar jets (the group of 5 around log L$_x^{\prime}$=42,
length=70kpc) are less secure since the jet emission is only of order
one resolution element from the quasar core emission for these
sources.

L$_x^{\prime}$ values are compromised by model dependency.  If quasar
jets were to come from synchrotron emission instead of IC/CMB
emission, the appropriate $\delta$ could well be of order 3 or 4
(similar to that for FRI's, and adequate to explain the jet
one-sidedness) instead of typical values like 10.  Thus the luminosity
correction when moving to the jet frame would be closer to a factor of
100 than 10,000 and the plot would be closer to a scaled version of
fig.~\ref{fig:obs_phys}.

For both of these figures we need to remember that 'low-power' and
'high-power' sources are so divided according to their total radio
luminosity.  When we plot the jet luminosity we are dealing with a
parameter that quantifies the jet loss, not the jet power.  Since FRI
jets are commonly thought of as being 'lossy', the underlying
assumption is that a larger fraction of the FRI jet power is radiated
than is the case for FRII jets.  Thus both the characteristic power
and the fractional energy lost to radiation for both classes of
sources are 'free' parameters and the resulting luminosities
(luminosity = total jet power $\times$ fractional loss to radiation)
would not necessarily be expected to be similar as in fig.~\ref{fig:lxp_phys}.

\section{Observations of Resolved Jets}\label{sec:data}

\subsection{Relevant Radio and Optical Considerations}\label{sec:3radopt}

In addition to the critical role of radio and optical flux densities,
which complement the X-ray intensities in defining the SED's of jet
knots, these longer wavelengths provide 2 critical capabilities for
jet observations: higher angular resolution than that of the CXO, and
polarization.  Moreover, in most cases, we are confident that we can
interpret the data on the basis of synchrotron emission rather than
being faced with the uncertainty of IC vs. synchrotron emission, as is
the case for the X-rays.

For the SED's, the IR-optical-UV data are usually those which
determine if a synchrotron spectrum (broken power law with high energy
cutoff) can be used to describe the radio to X-ray data.  For example,
\citet{jest05} find a spectral flattening from HST data in the 3C273
jet which is the basis for the claim that a simple synchrotron spectrum
cannot fit all the data.  Often the optical upper limit for a non
detection is used to preclude a single synchrotron component whereas
if no optical data were available, the radio and X-ray data could have
been interpreted as a single (broken) power law.

There are at least two observational problems affecting the
construction of SED's.  The first is the uncertainty that our
photometry is measuring the same entity in all bands.  The CXO
resolution is significantly worse than that of the HST so to gather
the counts for photometry, one needs at least a circle with radius of
0.5$^{\prime\prime}$.  Thus when measuring the SED of the knots in
e.g. 3C273 (fig.~\ref{fig:hst273}), we implicitly assume that the X-rays
are coming from the same emitting volume as the optical/radio, and not
from some additional volume such as a sheath around knots.
 
The second uncertainty is the absorption correction which mainly
affects the UV and soft X-ray data, and depends not only on the column
density to the source, but also on the gas to dust ratio.

\subsubsection{Morphology and Polarization at kpc scales}\label{sec:3kpc}

One of the more significant advances in understanding radio jets has
been achieved by \citet{lain02a,lain02b}.  For the case of FRI jets
for which both sides are visible and well resolved, they have been
able to use the laws of energy and momentum conservation to solve for
all the physical jet parameters assuming that the observed differences
in brightness and polarization between the two sides are caused by
relativistic effects only \citep[see][for a general discussion of
relativistic effects]{koni80}.  For 3C31 they find $\beta$=0.8 to 0.85
initially, then decelerating to $\beta$=0.2 at a few kpc's, with
loading by entrainment being the cause of the deceleration.  The
solution requires cross jet velocity structure: the outside has to be
going slower than the center.  This is consistent with, but does not
require, a simple 'spine/sheath' structure.

Optical and radio polarization have also been used to study the field
configuration in relation to the properties of internal shocks in
jets.  For example, \citet{perl05} find that the peaks in X-ray
brightness in the M87 jet coincide with minima of the optical
polarization.  They conclude that this is consistent with the location
of internal shocks which both produce the X-ray emission via particle
acceleration and change the magnetic field direction.  The observed
reduction in polarized signal would then be a result of beam smearing
over a region of swiftly changing field direction.


Other notable progress coming from optical data includes
\citet{bire99} who found features moving at $\approx$6c downstream
from M87/HST-1 (the same knot which later flared by a factor of 50
across the spectrum).  This demonstrates that at least mildly
relativistic velocities persist to kpc scales.  Several investigators
\citep[see for example][]{macc96} have also noted that optical
emission away from bright knots requires continuous acceleration
processes since the E$^2$ loss times are so short that the electrons
responsible for the observed emission cannot travel from the shock
locations; the same sort of argument was later deduced from similar
morphologies observed at X-ray frequencies.

\subsubsection{Parsec scale structures}\label{sec:3pc}

The most relevant aspect of VLBI work for X-ray jet physics is the
accumulating database containing monitoring of a reasonably large
sample of quasar, blazar, and BL Lac jets.  The original work was the
``2~cm~survey'' which has now become institutionalized on the web as
MOJAVE\footnote{http://www.physics.purdue.edu/astro/MOJAVE/}.  These
data provide a wealth of information such as the distribution of
beaming factors and jet velocities (if one accepts the notion that
observed proper motions of jet features reflect the underlying jet
velocity and that the sources are at distances indicated by their
redshifts).  \citet{kell04} find apparent velocities $\beta$ ranging
from zero to 15, with a tail extending up to 30 for individual
features.  With assumptions about brightness temperatures, this can be
translated to $\Gamma$ values covering a similar range.  If bulk
velocities of this magnitude persist to kpc scales, one of the
prerequisites of the IC/CMB model for X-ray jet emission will be
satisfied.

Another VLBA monitoring project is described in \citet{jors05}.  They
find a similar range for $\Gamma$ using intensity variability
timescales to estimate $\delta$, with most quasar components having
$\Gamma$ of order 16 to 18.  In both of these works, there is ample
evidence of non-ballistic motion: velocity vectors of components
mis-aligned with the jet vector.

\citet{gabu04} have used the transverse polarization
structure of jets resolved with VLBI to argue for a helical structure
for the magnetic field governing the emitting region; circumstantial evidence for
non-ballistic motions.  

\citet{ward98} argued for jet composition being a pair plasma based on
circular polarization inferences and \citet{hiro99} suggested that two
components in the jet of 3C279 were dominated by pair plasma on the
basis of electron density arguments.

\subsection{The X-ray Data}\label{sec:3xray}

We will not cover inferences from unresolved X-ray behavior of cores,
but concentrate on jet features for which we have some confidence that
the radio, optical, and X-ray emission comes from the same emitting
volume.  We make the usual assumptions that all relativistic plasmas
will emit both synchrotron and IC radiation and since most/all X-ray
jets are one sided, and these sides are the same as the those which
have VLBI superluminal jets, $\Gamma >$ 1, but not necessarily $\geq$
5 (for kpc scales).

\subsubsection{Jet Structure}\label{sec:3structure}

So far, there is very little transverse structure available from X-ray
data.  A notable exception is knot 3C120/k25 \citep{harr04} which is
resolved into 3 components. The jet of Cen A is well resolved since it
is the nearest jet source \citep{kraf02}. M87/knots A, B, and C are a
bit larger than the point spread function \citep{perl05} and Marshall
(private communication) reports that several features in the jet of
3C273 are also resolved by the CXO.

In general, there is good correspondence between jet knots mapped in
the radio, optical, and X-ray bands.  For both IC and synchrotron
emission models, this is expected in first-order approximations if
there is a single relativistic electron distribution responsible for
all observed emissions.  Relative intensities between bands can vary
depending on the relative magnitudes of energy densities in the
magnetic field and in the photons, as well as on the form of the
electron distribution, N(E), since different bands come from different
segments of N(E).

There are, however, a few cases of gross mis-alignment between an
X-ray feature and emissions at lower frequencies.  In the M87 jet,
beyond knot C (upper right of fig.~\ref{fig:m87}) the radio jet makes
a sudden excursion to the north downstream from a sharp gradient in
radio brightness (in the opposite sense to that of the leading edge of
knot A).  Although there is weak radio emission downstream of this
radio edge, the X-ray brightens.  One interpretation might be that the
radio jet encounters an 'obstacle' causing an internal shock and a jet
deflection.  The X-ray emission would then come from the obstacle, and
not properly be associated with the jet itself.  A similar situation
occurs just downstream of knot A in the jet of 3C273 where the radio
jet deflects to the South before resuming its principal direction,
whereas the X-ray, and in this case also the optical emission,
continues further North along the main jet vector defined by knot A
and the rest of the jet (fig.~\ref{fig:273K}).

\begin{figure}%
\caption{The first part of the jet in 3C273. The quasar itself is well
off this picture, up and to the left.  The false color image is from a
VLA map at 22 GHz, kindly supplied by R. Perley.  The radio beamsize
is 0.35$^{\prime\prime}$ FWHM.  The color scale is below the image and
given in Jy/beam.  The contours are from recent CXO data, smoothed
with a Gaussian of FWHM=0.25$^{\prime\prime}$.  The lowest contour is
0.008 ev s$^{-1}$ per 0.049$^{\prime\prime}$ pixel and successive contours
increase by factors of $\sqrt{2}$.  Note how the radio ridge line
heads directly south right after knot A; thus knot B1 lies on the
northern edge of the jet in the UV and X-ray, but on the southern side
of the jet in the radio.  Although B1 and B2 are not well resolved
with the CXO data, they are clearly separate in the radio and HST
images \citep{jest05}. }
\label{fig:273K}
\end{figure}

Another example of discrepant correspondence between radio and X-rays
is Cen A \citep{kraf02,hard03,hard04a}, shown in fig.~\ref{fig:cena}.
While many radio and X-ray features align well, albeit with quite
different relative intensities, there are a few X-ray knots that have
no obvious corresponding radio enhancements.  For PKS1136-135
\citep{samb02,samb04}, the radio emission associated with the first
bright X-ray knot ('A') is extremely weak; another example of the range of
relative intensities between radio and X-ray emissions.

\begin{figure}%
\caption{The jet in Cen A.  The colors show the X-ray image smoothed
with an 0.5$''$ FWHM Gaussian and the contours are from the VLA at 8
GHz with a beamsize of 1.1$^{\prime\prime}\times0.25^{\prime\prime}$
(PA of major axis $\approx$ 0$^\circ$).  The first contour level is
0.12 mJy/beam and successive intervals increase by factors of 4.  This
figure was provided by M. Hardcastle.}
\label{fig:cena}
\end{figure}

\paragraph{Offsets}\label{sec:3offsets}
      
Another common, although not universal, effect is the offset between
peak X-ray, optical, and radio brightness distributions when mapped
with similar angular resolutions.  It is generally the case that when
this occurs, the higher frequency brightness peaks at the upstream end
of the knot, and the underlying cause seems to be a steepening of the
spectra moving downstream \citep{hard03}.  A few examples are Cen A,
fig.~\ref{fig:cena} for knot A2 at RA=13h 25m 29s; M87 knots D and F
(fig.~\ref{fig:m87}); and knot B in PKS1127-145 (fig.~\ref{fig:1127}).
For the nearby sources Cen A and M87, the magnitude of the projected
offsets are of order tens of parsecs, whereas for PKS1127-145 at
z=1.18, the observed offset is of order 10 kpc.  Additional examples
from the FRI category are listed in \citet{bai03}, a paper devoted to
the offset effect.

\begin{figure}%
\caption{The jet of PKS1127-145.  The colors show a Chandra X-ray image
smoothed with an 0.5$''$ FWHM Gaussian and the contours are from the
VLA at 8 GHz with a beamsize of
0.78$^{\prime\prime}\times0.58^{\prime\prime}$ in PA=62$^\circ$.  The
first contour level is 0.2 mJy/beam and successive intervals increase
by factors of $\sqrt{2}$.  These new data will be published by
Siemiginowska et al. (2006, in preparation).}
\label{fig:1127}
\end{figure}

For the simplest synchrotron scenario, if electrons were to be
accelerated at a single location (i.e. a shock) and then be advected
down the jet, all synchrotron bands should coincide insofar as peak
brightness goes, even though downstream, we would expect to lose the
highest energy electrons sooner than the lower energy electrons
responsible for the radio emission.  This statement assumes perfect
angular resolution whereas normally, our beam sizes are not adequate
to discern these structural differences.  Thus even with the same
angular resolution, the peak brightness of the X-ray emission, being
centered on the shock, can occur upstream of the radio centroid which
has been shifted downstream a bit since the downstream plasma will
continue to produce radio emission (but not X-rays).

Another possibility to explain the observed offsets which is not
fine-tuned to the beam size is an increasing magnetic field strength
downstream from the shock, thereby enhancing the radio emissivity.
While the physical scales for the nearby sources are reasonably
consistent with these models (travel time matching E$^2$ halflives),
the 10kpc offset for PKS1127-145 (again assuming synchrotron emission)
would most likely rely on the second explanation.  It is also the
case, as emphasized earlier, that whereas we expect the electrons
responsible for the X-ray emission to travel no further than some tens
of light years, there is emission between the bright knots in many
jets, and this supports the presence of a distributed, quasi
continuous acceleration mechanism (see sec.~\ref{sec:1radi}).

For the IC/CMB model, currently there is no reasonable explanation as
to why the X-ray emission should drop off more rapidly than the
radio/optical synchrotron emission since the X-ray producing electrons
have $\gamma\leq$200, and thus much longer halflives than the
electrons responsible for the radio (and optical) emission.  There are
of course {\it ad hoc} possibilities such as a precursor shock
system (or some other mechanism) which would accelerate copious
numbers of electrons only up to some small value of $\gamma$ like
1000.

\paragraph{Progressions}\label{sec:3progressions}

Another effect which is closely associated with offsets is what we
call ``progressions''.  This term is applied to those jets for which
the X-ray intensity is highest at the upstream end, thereafter
generally decreasing down the jet, whereas the radio intensity
increases along the jet.  Progressions are rather common; in
\citet[fig. 5 of][]{samb04} this effect is shown for 7 quasars.  The
most striking example is 3C273 and profiles are shown in
fig.~\ref{fig:273prof}.  Note that the optical knots are of relatively
constant brightness.  If this jet were to be observed with a single
resolution element, there would be a clear offset between peak
brightnesses in the radio and X-ray which would most likely be
comparable to the length of the bright part of the jet, $\approx$
6$^{\prime\prime}$ (15 kpc).  Referring to fig.~\ref{fig:size}, we see
that the 3C273 jet is about the size of a single 'knot' in the
PKS1127-145 jet.  Thus we see that 'progressions' and 'offsets' can be
considered to be two manifestations of an underlying spectral
behavior.  Both offsets and progressions are observed in FRI and in
quasar jets although the common explanations for the two classes
differ.  For synchrotron models an increasing magnetic field strength
is posited, thereby increasing the synchrotron emissivity.  For IC
models, a gradual decrease in jet bulk velocity is assumed, leading to
a diminishing u'($\nu$) in the jet frame \citep{geor04}.

\begin{figure}*%
\caption{Profiles along the jet in 3C273.  The quasar is off the plots
  to the left.  The length shown is 16.7$^{\prime\prime}$ and the
  width used was 2$^{\prime\prime}$. Top panel: the X-ray data
  smoothed with a Gaussian of FWHM=0.25$^{\prime\prime}$.  Middle: a
  profile from an archival HST exposure (F622W), smoothed with a
  Gaussian of 0.5$^{\prime\prime}$.  Bottom: a profile from an
  8 GHz VLA map kindly supplied by R. Perley.  The clean beam is
  0.5$^{\prime\prime}$.  The vertical scales are linear, in arbitrary
  units.}
\label{fig:273prof}
\end{figure}

\paragraph{Emission between the knots}\label{sec:3interknot}

Although the conventional view of synchrotron jets is that electrons
responsible for X-ray emission cannot propagate more than a few light
years from their acceleration site, lower brightness emission is often
detected between the brighter knots.  In their survey of quasar jets,
\citet[][their sec. 4.3]{samb04} remark on this attribute for
PKS0605-085 and possibly also for 3C207 and PKS1136-135. Because of
the lower brightness levels in both radio and X-ray bands, it has been
difficult to obtain the data necessary to perform spectral tests for
the emission mechanism.  Quasi continuous jet emission is expected for
the IC/CMB process whereas a synchrotron hypothesis would require
continuous acceleration processes, as outlined in sec.~\ref{sec:1distrib}.

%

\subsubsection{Relative Intensities and SED's}\label{sec:3sed}

Combining radio, optical, and X-ray photometry to create a broad band
spectrum is the standard method to discriminate between synchrotron
and IC emission.  Resting on the common assumptions of a power law
distribution for the relativistic electrons and E$^2$ losses affecting
the highest energy electrons more severely than the lower energy
electrons, the notion that a one zone synchrotron source must have a
concave downward spectrum has been generally accepted.  This approach
can work even if only 3 flux densities are available (i.e. one radio,
one optical, and the X-ray measurement), but can become stronger with
more data, permitting estimates of the spectral index to be obtained
within each band.

There are several variations of this test such as plotting
$\alpha_{ro}$ against $\alpha_{ox}$ \citep[e.g.][fig. 4]{samb04}, or
simply demonstrating that an optical upper limit lies below the line
connecting the radio and X-ray data \citep[e.g.][]{schw06}.  Of the 34
quasars and FRII radio galaxies listed in Table~\ref{tab:fr2jets},
9 have been shown by \citet{samb04} to have knots with
$\alpha_{ro}>\alpha_{ox}$ and another 8 have optical upper limits
which preclude a simple synchrotron fit.  Twelve of the sources do not
yet have useful optical data available, and the remaining 5 consist of
a few FRII radio galaxies and a couple of quasars for which some knots
have spectra consistent with a synchrotron fit while others do not.

Although it is thus fairly simple to demonstrate that a simple
(i.e. 'single zone') synchrotron spectrum fails to apply to most knots
in quasar jets, once high quality data are available, serious problems
arise also for the single zone IC/CMB models.  Examples are
spine-sheath models devised to benefit both from high and low $\Gamma$
effects and 3C273 for which some knots seem to have
$\alpha_x~>~\alpha_r$, contrary to the expectation that electron
spectra will most likely flatten at low energies \citep{jest06}.

\subsubsection{Variability}\label{sec:3varib}

Since the CXO has been providing X-ray photometry of jet components
for only 6 years, to detect variability we require a small physical
size abetted by a significant value of $\delta$ to compress the
elapsed time in our frame.

Thus, clear intensity variability has so far only been found in Cen A
\citep{hard03} and M87 \citep{harr06a}.  This is not meant to preclude
the possibility of detection of variability in larger structures which
could well contain small scale structure.  If one were to ascribe the
factor of 100 difference in apparent luminosities between 3C273 and
M87 to a factor of 3 difference in $\delta$, then an event such as the
flare in knot HST-1 of M87 \citep{harr06a} would be easily detected.
The factor of 50 increase in the X-ray flux from HST-1 means that what
was once an inconspicuous X-ray knot for a time outshone the
remainder of the jet plus the unresolved core of M87.

Proper motion has been observed for radio features in Cen A
\citep{hard03} and optical features moving at up to 6c downstream of
HST-1 in M87 \citep{bire99}.  In both cases, the associated X-ray
features align with stationary radio or optical components.

\subsection{Jet Detection Statistics}\label{sec:statistics}

In the standard picture of AGN unification \citep{urry95}, there are
two main classes of AGN, low-power AGN (BL Lac objects and low-power
radio galaxies) and high-power AGN (Quasars and high-power radio
galaxies).  The differences of AGN within each class are explained
with a different degree of alignment between the line of sight and the
symmetry axis of the AGN (assumed to be parallel with the AGN jets).
BL Lac objects are interpreted to be the aligned versions of FRI radio
galaxies, and steep spectrum radio quasars (SSRQ - sources with a
radio spectral index $\alpha_{\rm r}$, $>$~0.5 at a few GHz) and flat
spectrum radio quasars (FSRQ, $\alpha_{\rm r}$, $<$~0.5) are
increasingly aligned versions of the FRII parent population.
\citet{urry95} derive the luminosity functions of the beamed AGN from
the luminosity function of the parent populations.  For their specific
source samples, their analysis indicates that FRII radio galaxies have
jets with bulk Lorentz factors of between 5 and 40, and that SSRQs and
FSRQs are FRIIs with jets aligned to within $\sim$38$^\circ$ and
$\sim$14$^\circ$ to the line of sight, respectively. The bulk Lorentz
factors of the FRI jets are less constrained but seem to be somewhat
lower than those of the FRII jets, and the jets of their radio
selected BL Lacs seem to be aligned to within 12$^\circ$ to the line
of sight.

In this context, we now consider the sources with X-ray jets (see
Tables 1 and 2).  The low-power sources with X-ray jets are mostly FRI
radio galaxies, except for 3C120, a Seyfert I galaxy and the two BL
Lacs PKS0521-365 and 3C371.  The high-power sources with X-ray jets
are all classified as quasars except for the four FRII radio galaxies
Pictor A, 3C219, 3C403 and PKS2152-69.  Remarkably, almost all X-ray
jets from the non-aligned FRI and FRII sources can be explained as
synchrotron emission from mildly relativistic jets; Lorentz factors of
a few are needed to explain the non-detection of counterjets. Most
sources for which the simple synchrotron picture does not work are
quasars. In this case, explaining the X-ray emission requires Lorentz
factors on the order of 10 and viewing angles on the order of
10$^\circ$.  The IC/CMB interpretation of the X-ray emission thus
indicates that the X-ray jets detected so far are similarly closely
aligned to the line of sight as the average radio-selected FSRQs used
in the FRII/quasar unification analysis described above.

\citet{samb04,mars05} used the CXO to study the fraction of sources
with X-ray jet emission for certain source samples. \citet{samb04} studied
sources with bright 1.4 GHz radio emission and a radio knot detection
more than 3$''$ away from the core.  Out of a sample of 17 sources,
X-ray jets were detected for 10 sources. \citet{mars05} studied two
samples of flat spectrum radio sources. One sample consisted of
sources selected for their high 5 GHz flux density. The other sample
consisted of sources with one-sided linear radio jet morphology. Out
of 19 sources of the first sample, 16 were detected with short CXO
observations. The detection probability in the second sample was
lower, but this finding was not statistically significant.

The samples used in these "survey-type studies" were
biased toward beamed sources. As the alignment of the sources 
is poorly constrained by the longer wavelength data, the high 
detection fraction with CXO cannot be used to argue for
or against the IC/CMB model for those sources for which simple
synchrotron models do not account for the X-ray emission.

As mentioned above, the radio spectral index can be used as an
indicator of the jet orientation relative to the line of sight. A
similar indicator is the lobe over core dominance at intermediate
radio frequencies like 5 GHz. A test of the IC/CMB model is to check
that the orientation parameters indicate an aligned jet for all
quasars that exhibit the bow-tie problem. Indeed only for one source
(PKS 1136-135) do we find at the same time $\alpha_r>>$0.5, lobe over
core dominance and a SED which indicates a bow-tie problem. However,
this source does not make a strong case against the IC/CMB model.
Only knot B exhibits the bow-tie problem and close inspection of the
radio-X-ray morphology shows that it may well be a hot-spot rather
than a jet knot.  We conclude that the IC/CMB scenario is not grossly
inconsistent with other orientation indicators.

\section{Discussion \& Summary}\label{sec:conclu}

\subsection{Critique of the Synchrotron Emission Model}\label{sec:4sync}

There seems to be little doubt that the X-ray emission from most or
all jets of FRI sources is dominated by the synchrotron process.  When
SED's are available, they are consistent with concave downward fits.
There are no problems with the synchrotron parameters such as magnetic
field strength or energy requirements.  Light curves for variable knots
also support the synchrotron model even if the predictions for
behavior at lower frequencies still need to be verified.  The
alternative of IC/CMB emission requires unreasonable beaming
parameters such as angles to the line of sight which are too small
compared to a host of other estimates.

Perhaps the most important implication to be deduced from FRI jets is
the necessity for distributed emission rather than a finite number of
shocks.  While we don't doubt the evidence for strong, discrete shocks
(e.g. a large gradient in radio brightness, often facing upstream),
some additional process is required.

The most likely alternatives for the 'additional process' are the
aforementioned 'distributed acceleration' and IC/CMB emission.  As
outlined in sec.~\ref{sec:1fluid}, one of the candidates for the
underlying jet 'medium' is electrons with $\gamma~<$~a few thousand.
If that option were to be correct, then it could well be the case that
even very modest values of $\Gamma, \delta$, and $\theta$ would
suffice for an IC/CMB model of inter-knot emission, and many of the
problems of this process for knots would not be present.  It is likely
that sufficient data have now accumulated in the relevant archives
that this test could be performed for a number of the brighter jets
with well defined knots.  

To explain jet segments devoid of detectable emission, this scenario
would indicate that jets are inherently intermittent.  Aside from
testing this suggestion by careful photometry, spectral analysis, and
calculation of beaming parameters, it would be a somewhat unbelievable
coincidence if the energy spectral index, p,
[N($\gamma$)$\propto~\gamma^{-p}$] were to be the same for the
postulated low energy electrons responsible for the jet's energy
transport and for the highest energy electrons with
$\gamma\approx~10^7$ responsible for the knot X-ray emission.  Thus
one could reasonably expect to see a marked change in $\alpha_x$
moving from knot to inter-knot regions.

The largest hurdle for the application of synchrotron emission to jets
of quasars comes from those cases for which the optical intensity is
so low (or undetected) that it precludes a concave downward spectral
fit from radio to X-rays.  The associated 'bow-tie' problem
(sec.~\ref{sec:1bowtie}) has been reported also for FRI sources
\citep[e.g. 3C120, knot `k25'][]{harr04}.  None of the possible
solutions has been accepted by the community and progress on this
issue depends on a demonstration that a key ingredient of spectral
hardening at high energy is indicated by some independent means.
Examples would be the confirmation of a prediction from the two-zone
model or finding independent support for the shear layer acceleration
model.

\subsection{Critique of the IC/CMB Emission Model}\label{sec:4ic}

The idea of augmenting u($\nu$) compared to u(B) in the jet
frame had been used for jets close to black holes where u($\nu$) was
thought to be dominated by UV radiation \citep[e.g.][]{derm94,blan95,siko97}.
\citet{celo01} and \citet{tave00} applied this concept to kpc scale jets for
which the CMB dominates u($\nu$).  By positing that the X-ray knots of
PKS0637-752 had a value of $\Gamma\approx$10, similar to the values
deduced from superluminal proper motions for the pc scale jet
\citep{ting00}, they were able to show that IC/CMB was able to explain
the observed X-ray intensities while still maintaining equipartition
conditions between u(B) and u(p) (where u(p) is the energy density in
relativistic particles).  This idea was quickly adopted by the
community because it was already realized that the preponderance of
one-sided X-ray jets requires $\Gamma\geq$3 or 4 and it provided a
solution to the vexing problem of too little optical intensity to
provide a reasonable synchrotron fit to the spectrum.

Additional support for this model is supplied by the jets that show
the progression, discussed above, with a decreasing ratio of X-ray to
radio intensity moving away from the core.  Under the IC/CMB model,
all that is required is a general deceleration of the jet, thereby
reducing u$^{\prime}$($\nu$) in the jet frame \citep[e.g. for
3C273,][]{samb01}.  There is, of course the problem of explaining why
the IC/CMB X-ray intensity of 3C273/knot A happens to fall so close to
the extrapolation of the radio/optical synchrotron spectrum
\citep{mars01}.

There are a number of additional uncertainties and problems for the
IC/CMB model although none of these represents a definitive
refutation.

\paragraph{Offsets and lifetime considerations}\label{sec:4offsets}

In sec.~\ref{sec:3offsets} we discussed offsets between X-ray and
radio brightness distributions of jet knots.  The low energy
($\gamma\approx$100) electrons responsible for the X-ray emission will
have E$^2$ lifetimes in excess of 10$^6$ yr; sufficient to travel to
the end of even a Mpc jet.  Thus when confronted with a knotty X-ray
jet, the question arises: once a copious supply of these electrons are
generated (e.g. at knot A in 3C273), why does the emission fade to a
low level and then rise again for the next knot instead of forming a
continuous or cumulatively brightening jet?  One might devise a rather
contrived scenario by having the beaming factor decrease to end one
knot, and then either increase at the location of the next knot, or
posit the injection of enough new electrons to produce a bright knot,
even if the beaming factor were smaller than that of the first knot.
This explanation is unsatisfactory when the radio emission is
considered since it should follow the X-ray behavior if $\delta$ is
the controlling factor.  The same sort of problem affects the observed
offsets \citep{staw04a,atoy04}; the X-ray brightness should persist
further downstream than the optical and radio, the opposite of what is
observed.

\paragraph{Energetics}\label{sec:4energy}

\citet{derm04,atoy04} have made a comprehensive review of the X-ray
emission processes and emphasize that the original formulation of
IC/CMB emission worked on the equipartition assumption based on the
radio data.  When the electron spectrum is extended down to the low
energies required by IC/CMB, the particle energy density [and hence
also u(B)] increase significantly.  This leads them to conclude that
excessive energies for the jet are required, even under the
``optimistic'' assumption that the jet is made of an electron/positron
plasma without cold protons.  For PKS0637-752, they find
the kinetic luminosity is $\geq7\times10^{46}$~erg~s$^{-1}$ for
$\delta$=27, $\theta\leq2^{\circ}$, and increases for more reasonable
beaming parameters.  The associated total energy is $\geq10^{57}$ ergs
for the case of $\delta=\Gamma\approx$10, $\theta\approx$5$^{\circ}$.

\paragraph{Uncertainty of extrapolation of the electron spectrum}\label{sec:4extrap}

One of the implicit assumptions of every IC/CMB calculation (i.e. to
determine the required beaming parameters to explain the observed
radio and X-ray intensities) is that the electron spectrum extends to
very low energies with a slope p=2$\alpha_r$+1.  If that were to be
the case, then $\alpha_x$ should have the same value as $\alpha_r$.
However, as demonstrated in fig.~\ref{fig:extrap}, we currently have
no knowledge that this condition holds.  If $\alpha_x$ is less than
$\alpha_r$, it would indicate a low frequency break to a flatter
spectrum and the estimated beaming parameters would be wrong.  With
fewer low energy electrons than assumed by the extrapolation, $\Gamma$
and $\delta$ would have to be larger and $\theta$ correspondingly
smaller, exacerbating some of the problems listed above.  It is, of course, 
conceivable that the electron spectrum takes an upturn at
low energies, in which case the error goes in the opposite direction.

Another assumption often, but not always present is that of
equipartition.  Since every calculation requires a value of the
magnetic field in order to move from the observed segment of the
synchrotron spectrum to obtain the corresponding segment of the
electron spectrum, the usual method is to assume equipartition.  When
that constraint is removed as in the case of arguing for a field
strength well below equipartition \citep[e.g.][]{kata05}, the electron
spectrum can be considered to be undefined and one can conjure up
whatever number of low energy electrons are needed to explain the X-rays
for a given beaming factor.  In the case of \citet{kata05}, a small
value of $\Gamma$ was invoked based on radio asymmetry arguments
\citep{ward97}.  The initial analysis of PKS0637-752 \citep{schw00} also
suggested substantial dominance of u(p) since the IC/CMB scenario with
beaming was not widely known at that time.

Finally, not only do IC/CMB models require a substantial extrapolation
of the electron energy distribution to low energies with a fixed power
law, they also require some fine tuning of a strict cutoff in the
distribution at some slightly lower $\gamma$ in order not to
over-produce the optical emission \citep[e.g.][Table 7]{samb04}.

\paragraph{Small angles to the line of sight and physical length of jets}\label{sec:4theta}

From recent quasar surveys with the CXO \citep{samb04,mars05,schw06},
fitting IC/CMB models yield $\delta$ values which range from 3 to 11. For
$\Gamma=\delta$, this means $\theta$ is most commonly between
4$^{\circ}$ and 11$^{\circ}$.  Since most X-ray jets are
reasonably straight, the physical length of jets sometimes exceeds 1
Mpc (n.b. the jet lengths given in Tables~\ref{tab:fr1jets} and
\ref{tab:fr2jets} refer primarily to the X-ray extent; the radio jet
is often longer).

Many workers \citep[e.g.][]{derm04} find Mpc scale quasar jets
uncomfortably long, and it is certainly the case that most FRII radio
galaxies, the 'face-on' counterparts of quasars under the unified
scheme, are much smaller.  However, there are a small number of 'giant
radio galaxies', and even a few quasars with sizes considerably
greater than 1 Mpc \citep[e.g.][]{rile90}.

In some cases source morphology inferences are in conflict with small
$\theta$.  \citet{wils01} argue that if the IC/CMB model with
equipartition is applied to the jet in Pictor A, $\Gamma=\delta$=7.2
and $\theta$=8$^{\circ}$.  Such an angle to the line of sight would
mean that the total extent of the source would be on the order of 3
Mpc and the hotspots at the outer end of each lobe, should be seen
projected onto the radio lobes instead of protruding beyond the lobes
as they are actually situated.  Although \citet{wils01} conclude that
an IC/CMB model at a more reasonable $\theta\approx23^{\circ}$ would
require B$<$B$_{eq}$, \citet{hard05b} subsequently have made a strong
case that the X-ray emission from the Pictor A jet is synchrotron
emission, not IC/CMB.

\paragraph{Expectations for jets with z$>$1}\label{sec:4highz}

\citet{schw02} has argued that at higher redshifts there should be
more jet detections since the increase in u($\nu$) of the CMB by the
factor (1+z)$^4$ will compensate for the usual redshift dimming of
surface brightness.  In addition to this effect, we might expect to
see more of the lower $\Gamma$ jets with larger beaming cones because
the (1+z)$^4$ factor already will statistically increase the ratio of
u($\nu$)/u(B) regardless of the $\Gamma^2$ factor from the jet's bulk
velocity.  So far, these predictions have not been realized
\citep{bass04}, and \citet[][see their fig. 10]{kata05} have
emphasized that the required $\delta$ values for the IC/CMB model generally
decrease with redshift.  At this stage, the only quasar jet detection with 
z substantially $\geq$2 is GB 1508+5714 (z=4.3).

\subsection{Tests to differentiate between synchrotron and IC/CMB models}\label{sec:4test}

The basic tenet of the IC/CMB model for jet knots is that the X-ray
emission is sampling the low energy end of the power law electron
distribution.  Therefore, the IC emission must continue to higher
frequencies, unlike the synchrotron spectrum which is already
relatively steep, and most likely will show an exponential cutoff at
somewhat harder X-ray energies than available with the CXO.  If we
could measure the X-ray spectrum of quasar knots at much higher
frequencies, and found a smooth continuation, it would be a clear
confirmation of the IC/CMB model.  If on the other hand, we were to
find a cutoff in the X-ray spectrum, that would indicate synchrotron
emission.  Unfortunately, there are no real prospects of convincingly
performing this test since it is so difficult to reach the required
sensitivity and angular resolution above 10 keV. The CXO band is too
narrow to define the expected cutoff which may well be smeared over a
wide frequency band by internal source structure.

Another option for discriminating these emission mechanisms will
become available as new radio telescopes with unprecedented
sensitivity and resolution at low frequencies come on line in the next
few years.  Both LOFAR in the Netherlands and the LWA (Long Wavelength
Array) in the US will have the capability to resolve jet knots and
determine the characteristics of the electron distribution at the low
energies of interest (fig.~\ref{fig:extrap}).  Each of these
instruments will have a reasonably wide frequency coverage so that not
only the amplitude, but also the slope of the low frequency emission
can be measured.  If we find that the low frequency radio data
indicates that the spectrum flattens significantly or has a low
frequency cutoff, then the IC/CMB model will have serious problems.

Optical and IR telescopes can be used to achieve detections of jet
knots which currently have only upper limits.  This band plays a
crucial role for the IC/CMB model because there is still substantial
uncertainty as to the origin of the currently detected optical
features: is this emission from the top end of the synchrotron
spectrum or the bottom end of the IC spectrum?  Robust detections and
photometry at several wavelengths should clarify this problem which
impacts on the general 'fine tuning' of the low energy end of the
electron spectrum (sec.~\ref{sec:4extrap}).

\subsection{Detectability of the Extended Jet Emission By Gamma-Ray Telescopes}

The {\it EGRET} detector on board the Compton Gamma-Ray Observatory
established that blazars, AGNs with their jets aligned with the line
of sight, are strong sources of gamma-rays. The EGRET experiment
(approximately 20 MeV to 30 GeV, or 5$\times10^{21}$ to
7$\times10^{24}$~Hz) detected a total of 66 blazars with redshifts up
to $z$ $\sim$ 2 \citep{hart99}. A small number of blazars (currently
10) with redshifts between 0.031 and 0.186 have been detected at even
higher energies (GeV to TeV, frequencies above 10$^{25}$~Hz) with
ground-based Cherenkov telescopes \citep{kraw05}.  Rapid gamma-ray
flux variability on time scales between 15 min and a few hours,
together with assumptions about infrared to UV emission co-spatially
emitted with the gamma-rays, have been used to derive a lower limit on
the Doppler factor $\delta ^{>}_\sim$ 10 of the emitting plasma based
on gamma-ray opacity arguments \citep{gaid96,matt97}.  All of these
observations refer to very small physical scales, resulting in
completely unresolved data from the nuclear regions.

If the extended jet emission detected by Chandra indeed originates
from the IC/CMB process, the IC component should in principle be
detectable in the MeV/GeV energy range with the Gamma-ray Large Area
Space Telescope (GLAST) to be launched in 2007 \citep{mcen04}, and
possibly also in the GeV/TeV energy regime with ground-based
telescopes like H.E.S.S., VERITAS, MAGIC, and CANGAROO III
\citep{ahar04,week03}.  GLAST has a sensitivity for the flux above 100
MeV of 3$\times~10^{-13}$~ergs~cm$^{-2}$s$^{-1}$ for 5 yrs of
sky-survey observations.  Cherenkov telescopes like VERITAS and
H.E.S.S. have a 100 GeV sensitivity of 9$\times~10^{-13}$
ergs~cm$^{-2}$s$^{-1}$ for 100hrs integration. These estimates have
been derived by the instrument teams for photon indices of 2.  For
harder photon spectra with indices of 1.5, the $\nu\times f_{\nu}$
sensitivities are about a factor of 2 better.  IC/CMB models predict
gamma-ray fluxes between 10$^{-13}$ and a few times 10$^{-12}$ ergs cm$^{-2}$
s$^{-1}$ \citep{tave04a,derm04} so these new observatories should have
sufficient sensitivity for detection.

The angular resolution of GLAST for a single photon will be 3.4$^\circ$ at 100
MeV, and 0.1$^\circ$ at 10 GeV; typical source localization accuracies
will be tens of arc-minutes near detection threshold and 0.5 arcmin
for very strong sources
\footnote{http://www-glast.slac.stanford.edu/}.  For most sources, the
angular distance between the core and kpc-scale jet is only a few
arcsec, and GLAST will not be able to distinguish between core and jet
emission on the basis of the spatial information.  Furthermore,
variability studies will be limited to rather long time scales and
large fractional flux variations.

Cherenkov telescopes have better angular resolutions
($\approx$0.1$^\circ$) and source localization accuracies
($\approx$20$''$).  For $\sim$100 GeV photons however, the transparency
of the Universe is limited to redshifts on the order of 0.5 owing to
the gamma-rays pair-producing on IR background photons (1 to 40
microns) from galaxies.  Detection and identification of gamma-rays
from kpc-scale jets would thus require very strong sources with very
extended X-ray jets at low redshifts; the chances for obtaining
unambiguous results are not promising.

\subsection{Prospects}\label{sec:prospects}

\subsubsection{Synchrotron Emission}

In general, synchrotron emission is a powerful diagnostic of
relativistic plasmas, and in the particular case of X-ray
frequencies, informs us as to the location of acceleration sites.  The
major problem is the unknown magnetic field strength which precludes a
direct determination of the electron energy distribution.

Since the X-ray emitting electrons have such a high energy and
consequently short lifetime, we expect variability in jets will
continue to offer new insights.  With a multifrequency monitoring, it
should be possible to disentangle light travel times from E$^2$
halflives and thus obtain a different estimate of the magnetic field
strength and or u'($\nu$) as well as $\delta$ \citep{harr06a}.

As more jets are studied with greater sensitivity, we believe the
chances are good that we should find a few objects that display the
effects of a high energy cutoff in the CXO band.  Though we assume that
all synchrotron plasmas have cutoffs, few if any have actually been
observed in radio, optical, or X-rays.  This result would impact the
acceleration scenario by providing an estimate of the extent in energy
of the electron distribution.

On the theoretical front, we need additional ideas of how deviations
from a power law electron spectrum can occur.  The two proposals
currently available are rather restricted in applicability and should
be further developed.

\subsubsection{IC emission}

If the jet X-ray emission from powerful sources is indeed from the
IC/CMB process, we can study different attributes of the underlying
relativistic plasma than those involved in synchrotron emission.  In
particular, we can obtain vital information about the low energy part
of the electron spectrum.  Both the amplitude and slope for
$\gamma\leq$1000 are germane to the injection problem for shock
acceleration as well as permitting greatly improved estimates of the
total particle energy density and hence the energetics of the emitting
plasma.

As is well known, estimates of the photon energy density are amenable
to direct observational input, and this permits us to pass more
confidently from the emission spectrum to the electron spectrum.  Once
the electron spectrum is known, then the observed synchrotron
component will provide the magnetic field strength.  The
basic physics is understood and IC emission is mandatory in all
relativistic plasmas.  The only questions are, how much emission is
there and what is the frequency range of the emission.

For the beaming IC/CMB model applied to jets, some 'paradigm shifts'
will be in order.  If current estimates of beaming parameters are
correct, many of the relatively bright X-ray knots are, in their own
frame, rather unimpressive: luminosities of order 10$^{38}$ to
10$^{39}$erg~s$^{-1}$ would be common and the canonical
10$^{44}$erg~s$^{-1}$ would no longer be relevant.

Another effect means that our view of jets close to the line of sight
is actually a stretched out version of the time history of a very
small fraction of the 'current jet length' (by which we mean the
distance from the outermost knot or hotspot to the core, at the time
we observe the jet tip).  This can be quickly grasped by reversing
time and sending a signal from the earth to the quasar.  As the
wavefront of our signal passes the jet tip, the jet is moving
relativistically towards the quasar.  For example, take a 100,000
l.y. jet at 5$^{\circ}$ to our line of sight.  If the jet has a bulk
velocity of 0.99c, by the time our wavefront reaches the quasar about
98.6\% of the jet (as it existed when our wavefront first reached the
tip) has now been swallowed by the black hole, and is thus not
observable by us.  What we see, which appears to be 100,000 l.y. in
length, is actually just the 1,400 l.y. long tip of the 'current jet',
as it was at progressively earlier times as we move back from the tip.
The most important aspect of this effect is to make the necessary
adjustments when comparing quasar jets to those lying closer to the
plane of the sky.  What might we actually be studying if all we see is
1\% of the 'current jet length?  The hotspot?  If so, what we call
knots in the jets would actually be bits of the hotspot brightening
and fading over its 100,000 year long journey to its 'present'
location.

\subsection{Summary}\label{sec:4summary}

Within a few years, the uncertainty as to the X-ray emission process
for quasar jets should be eliminated and then we will either have a
method of measuring the low energy end of the relativistic electron
distribution (if the IC/CMB model applies) or we will have new
insights into the behavior and loss mechanisms affecting the highest
energy electrons (if synchrotron models apply).  If we are convinced
that the IC/CMB model is correct, then a number of conclusions are
already clear: detected quasar jets lie close to the line of sight and
have large Lorentz factors.  That in turn means we can solve for some
of the basic jet parameters such as energy flux, and most likely we
will improve our understanding of cross-jet velocity structure: many
different lines of argument point to the necessity of some sort of
'spine-sheath' structure.

In part 2 of this review, we have examined the differences between the
jets of FRI radio galaxies and those of quasars.  Will the
distinctions in jet length and luminosity translate to differences in
X-ray emission process?  If so, why are there so many similarities
between low power and high power sources such as offsets and
progressions?  We may also expect to better understand the underlying
reasons for brightness fluctuations along jets, and if the small knots
of FRI jets have the same genesis as the kpc scale knots in quasar
jets.  All of these lines of investigation will hopefully elucidate
the dichotomy between the plasma that emits the radiation we observe
and the medium which transports the energy and momentum over such vast
distances.

\section{ACKNOWLEDGMENTS}

We thank C. Cheung, S. Jester, M. Hardcastle, and many other
colleagues for useful discussions.  C. Cheung, and L. Stawarz kindly
gave us helpful comments on the manuscript and the editor,
R. Blandford, provided valuable advice.  This work has made use of
NASA's Astrophysics Data System Bibliographic Services and the XJET
website.  Partial support was provided by NASA contract NAS8-03060 and
grant GO3-4124A.  HK thanks the Department of Energy for support in
the framework of the Outstanding Junior Investigator program.


\small
\bibliography{master}

\begin{thebibliography}{}
\expandafter\ifx\csname natexlab\endcsname\relax\def\natexlab#1{#1}\fi

\bibitem[{{Aharonian}(2002)}]{ahar02}
{Aharonian} FA. 2002.
\newblock \textit{\mnras} 332:215--230

\bibitem[{{Aharonian}(2004)}]{ahar04}
{Aharonian} FA. 2004.
\newblock \textit{Very high energy cosmic gamma radiation: a crucial window on
  the extreme universe}.
\newblock River Edge, NJ: World Scientific Publishing

\bibitem[{Aldcroft et~al.(2003)Aldcroft, Siemiginowska, Elvis, Mathur, Nicastro
  \& Murray}]{aldc03}
Aldcroft T, Siemiginowska A, Elvis M, Mathur S, Nicastro F, Murray S. 2003.
\newblock \textit{\apj} 597:751

\bibitem[{{Atoyan} \& {Dermer}(2004)}]{atoy04}
{Atoyan} A, {Dermer} CD. 2004.
\newblock \textit{\apj} 613:151--158

\bibitem[{Bahcall et~al.(1995)Bahcall, Kirhakos, Schneider, Davis, Muxlow
  et~al.}]{bahc95}
Bahcall JN, Kirhakos S, Schneider DP, Davis RJ, Muxlow TWB, et~al. 1995.
\newblock \textit{\apjl} 452:L91

\bibitem[{{Bai} \& {Lee}(2003)}]{bai03}
{Bai} JM, {Lee} MG. 2003.
\newblock \textit{\apjl} 585:L113--L116

\bibitem[{{Bassett} et~al.(2004){Bassett}, {Brandt}, {Schneider}, {Vignali},
  {Chartas} \& {Garmire}}]{bass04}
{Bassett} LC, {Brandt} WN, {Schneider} DP, {Vignali} C, {Chartas} G, {Garmire}
  GP. 2004.
\newblock \textit{\aj} 128:523--533

\bibitem[{Begelman, Blandford \& Rees(1984)}]{bege84}
Begelman MC, Blandford RD, Rees MJ. 1984.
\newblock \textit{RvMP} 56:255

\bibitem[{Bell(1978)}]{bell78}
Bell AR. 1978.
\newblock \textit{MNRAS} 182:147

\bibitem[{{Beresnyak}, {Istomin} \& {Pariev}(2003)}]{bere03}
{Beresnyak} AR, {Istomin} YN, {Pariev} VI. 2003.
\newblock \textit{\aap} 403:793--804

\bibitem[{{Bicknell}(1995)}]{bick95}
{Bicknell} GV. 1995.
\newblock \textit{\apjs} 101:29

\bibitem[{{Bicknell} \& {Begelman}(1996)}]{bick96}
{Bicknell} GV, {Begelman} MC. 1996.
\newblock \textit{\apj} 467:597

\bibitem[{Biretta, Sparks \& Macchetto(1999)}]{bire99}
Biretta JA, Sparks WB, Macchetto F. 1999.
\newblock \textit{\apj} 520:621

\bibitem[{{Birkinshaw}, {Worrall} \& {Hardcastle}(2002)}]{birk02}
{Birkinshaw} M, {Worrall} DM, {Hardcastle} MJ. 2002.
\newblock \textit{\mnras} 335:142--150

\bibitem[{Blandford \& Payne(1982)}]{blan82}
Blandford R, Payne DG. 1982.
\newblock \textit{MNRAS} 199:883

\bibitem[{Blandford \& Rees(1974)}]{blan74}
Blandford R, Rees M. 1974.
\newblock \textit{MNRAS} 169:395

\bibitem[{{Blandford}(1976)}]{blan76}
{Blandford} RD. 1976.
\newblock \textit{\mnras} 176:465--481

\bibitem[{Blandford \& Levinson(1995)}]{blan95}
Blandford RD, Levinson A. 1995.
\newblock \textit{\apj} 441:79

\bibitem[{Blandford \& Ostriker(1978)}]{blan78}
Blandford RD, Ostriker JP. 1978.
\newblock \textit{\apj} 221:L29

\bibitem[{{Blandford} \& {Znajek}(1977)}]{blan77}
{Blandford} RD, {Znajek} RL. 1977.
\newblock \textit{\mnras} 179:433--456

\bibitem[{{Bodo} et~al.(2003){Bodo}, {Rossi}, {Mignone}, {Massaglia} \&
  {Ferrari}}]{bodo03}
{Bodo} G, {Rossi} P, {Mignone} A, {Massaglia} S, {Ferrari} A. 2003.
\newblock \textit{New Astronomy Review} 47:557--559

\bibitem[{{Bridle} et~al.(1994){Bridle}, {Hough}, {Lonsdale}, {Burns} \&
  {Laing}}]{brid94}
{Bridle} AH, {Hough} DH, {Lonsdale} CJ, {Burns} JO, {Laing} RA. 1994.
\newblock \textit{\aj} 108:766--820

\bibitem[{Brunetti et~al.(2003)Brunetti, Harris, Sambruna \& Setti}]{brun03}
Brunetti G, Harris D, Sambruna R, Setti G. 2003.
\newblock \textit{NewAR} 47:411--712

\bibitem[{{Celotti} \& {Fabian}(1993)}]{celo93}
{Celotti} A, {Fabian} AC. 1993.
\newblock \textit{\mnras} 264:228

\bibitem[{{Celotti}, {Ghisellini} \& {Chiaberge}(2001)}]{celo01}
{Celotti} A, {Ghisellini} G, {Chiaberge} M. 2001.
\newblock \textit{\mnras} 321:L1--L5

\bibitem[{{Chartas} et~al.(2002){Chartas}, {Gupta}, {Garmire}, {Jones}, {Falco}
  et~al.}]{char02}
{Chartas} G, {Gupta} V, {Garmire} G, {Jones} C, {Falco} EE, et~al. 2002.
\newblock \textit{\apj} 565:96--104

\bibitem[{{Chartas} et~al.(2000){Chartas}, {Worrall}, {Birkinshaw},
  {Cresitello-Dittmar}, {Cui} et~al.}]{char00}
{Chartas} G, {Worrall} DM, {Birkinshaw} M, {Cresitello-Dittmar} M, {Cui} W,
  et~al. 2000.
\newblock \textit{\apj} 542:655--666

\bibitem[{{Cheung}(2004)}]{cheu04}
{Cheung} CC. 2004.
\newblock \textit{\apjl} 600:L23--L26

\bibitem[{{Chiaberge} et~al.(2000){Chiaberge}, {Celotti}, {Capetti} \&
  {Ghisellini}}]{chia00}
{Chiaberge} M, {Celotti} A, {Capetti} A, {Ghisellini} G. 2000.
\newblock \textit{\aap} 358:104--112

\bibitem[{Chiaberge et~al.(2003)Chiaberge, Gilli, Macchetto, Sparks \&
  Capetti}]{chia03}
Chiaberge M, Gilli R, Macchetto F, Sparks W, Capetti A. 2003.
\newblock \textit{\apj} 582:645

\bibitem[{Comastri et~al.(2003)Comastri, Brunetti, Dallacasa, Bondi, Pedani \&
  Setti}]{coma03}
Comastri A, Brunetti G, Dallacasa D, Bondi M, Pedani M, Setti G. 2003.
\newblock \textit{MNRAS} 340:L52

\bibitem[{Coppi(1999)}]{copp99}
Coppi PS. 1999.
\newblock In \textit{Relativistic\,Jets\,in\,AGNs}, eds. M~Ostrowski, M~Sikora,
  G~Madejski, M~Begelman. Jagellonian\,University\,Press.
\newblock Astro-ph/9903162

\bibitem[{Curtis(1918)}]{curt18}
Curtis HD. 1918.
\newblock \textit{Pub. Lick. Obs.} 13:31

\bibitem[{Dennett-Thorpe et~al.(1997)Dennett-Thorpe, Bridle, Scheuer, Laing \&
  Leahy}]{denn97}
Dennett-Thorpe J, Bridle A, Scheuer P, Laing R, Leahy J. 1997.
\newblock \textit{MNRAS} 289:753

\bibitem[{Dermer \& Atoyan(2004)}]{derm04}
Dermer CD, Atoyan A. 2004.
\newblock \textit{\apj} 611:L9

\bibitem[{{Dermer} \& {Atoyan}(2002)}]{derm02}
{Dermer} CD, {Atoyan} AM. 2002.
\newblock \textit{\apjl} 568:L81--L84

\bibitem[{{Dermer} \& {Schlickeiser}(1994)}]{derm94}
{Dermer} CD, {Schlickeiser} R. 1994.
\newblock \textit{\apjs} 90:945--948

\bibitem[{{Evans} et~al.(2005){Evans}, {Hardcastle}, {Croston}, {Worrall} \&
  {Birkinshaw}}]{evan05}
{Evans} DA, {Hardcastle} MJ, {Croston} JH, {Worrall} DM, {Birkinshaw} M. 2005.
\newblock \textit{\mnras} 359:363--382

\bibitem[{Fabian, Celotti \& Johnstone(2003)}]{fabi03}
Fabian A, Celotti A, Johnstone R. 2003.
\newblock \textit{MNRAS} 338:L7

\bibitem[{{Fukue}, {Tojyo} \& {Hirai}(2001)}]{fuku01}
{Fukue} J, {Tojyo} M, {Hirai} Y. 2001.
\newblock \textit{\pasj} 53:555--563

\bibitem[{{Gabuzda}, {Murray} \& {Cronin}(2004)}]{gabu04}
{Gabuzda} DC, {Murray} {\' E}, {Cronin} P. 2004.
\newblock \textit{\mnras} 351:L89--L93

\bibitem[{Gaidos et~al.(1996)Gaidos, Akerlof, Biller \& et~al.}]{gaid96}
Gaidos JA, Akerlof CW, Biller SD, et~al. 1996.
\newblock \textit{Nature} 383:319

\bibitem[{{Georganopoulos} \& {Kazanas}(2003)}]{geor03}
{Georganopoulos} M, {Kazanas} D. 2003.
\newblock \textit{\apjl} 589:L5--L8

\bibitem[{{Georganopoulos} \& {Kazanas}(2004)}]{geor04}
{Georganopoulos} M, {Kazanas} D. 2004.
\newblock \textit{\apjl} 604:L81--L84

\bibitem[{{Ghisellini}, {Tavecchio} \& {Chiaberge}(2005)}]{ghis05}
{Ghisellini} G, {Tavecchio} F, {Chiaberge} M. 2005.
\newblock \textit{\aap} 432:401--410

\bibitem[{Graziani, Lamb \& Donaghy(2006)}]{graz06}
Graziani C, Lamb DQ, Donaghy TQ. 2006.
\newblock \textit{\apj} submitted [astro-ph/0505623]

\bibitem[{Hardcastle, Birkinshaw \& Worrall(2001)}]{hard01}
Hardcastle M, Birkinshaw M, Worrall D. 2001.
\newblock \textit{MNRAS} 326:1499

\bibitem[{Hardcastle \& Croston(2005)}]{hard05b}
Hardcastle M, Croston J. 2005.
\newblock \textit{MNRAS} 363:649

\bibitem[{{Hardcastle} et~al.(2004{\natexlab{a}}){Hardcastle}, {Harris},
  {Worrall} \& {Birkinshaw}}]{hard04b}
{Hardcastle} MJ, {Harris} DE, {Worrall} DM, {Birkinshaw} M. 2004{\natexlab{a}}.
\newblock \textit{\apj} 612:729--748

\bibitem[{{Hardcastle} et~al.(2002){Hardcastle}, {Worrall}, {Birkinshaw},
  {Laing} \& {Bridle}}]{hard02}
{Hardcastle} MJ, {Worrall} DM, {Birkinshaw} M, {Laing} RA, {Bridle} AH. 2002.
\newblock \textit{\mnras} 334:182--192

\bibitem[{{Hardcastle} et~al.(2005){Hardcastle}, {Worrall}, {Birkinshaw},
  {Laing} \& {Bridle}}]{hard05a}
{Hardcastle} MJ, {Worrall} DM, {Birkinshaw} M, {Laing} RA, {Bridle} AH. 2005.
\newblock \textit{\mnras} 358:843--850

\bibitem[{{Hardcastle} et~al.(2003){Hardcastle}, {Worrall}, {Kraft}, {Forman},
  {Jones} \& {Murray}}]{hard03}
{Hardcastle} MJ, {Worrall} DM, {Kraft} RP, {Forman} WR, {Jones} C, {Murray} SS.
  2003.
\newblock \textit{\apj} 593:169--183

\bibitem[{{Hardcastle} et~al.(2004{\natexlab{b}}){Hardcastle}, {Worrall},
  {Kraft}, {Forman}, {Jones} \& {Murray}}]{hard04a}
{Hardcastle} MJ, {Worrall} DM, {Kraft} RP, {Forman} WR, {Jones} C, {Murray} SS.
  2004{\natexlab{b}}.
\newblock \textit{Nuclear Physics B Proceedings Supplements} 132:116--121

\bibitem[{{Hardee}, {Walker} \& {G{\'o}mez}(2005)}]{hardee05}
{Hardee} PE, {Walker} RC, {G{\'o}mez} JL. 2005.
\newblock \textit{\apj} 620:646--664

\bibitem[{Harris, Mossman \& Walker(2004)}]{harr04}
Harris D, Mossman A, Walker R. 2004.
\newblock \textit{\apj} 615:161

\bibitem[{{Harris}, {Carilli} \& {Perley}(1994)}]{harr94}
{Harris} DE, {Carilli} CL, {Perley} RA. 1994.
\newblock \textit{\nat} 367:713

\bibitem[{Harris et~al.(2006)Harris, Cheung, Biretta, Junor, Perlman
  et~al.}]{harr06a}
Harris DE, Cheung CC, Biretta JA, Junor W, Perlman ES, et~al. 2006.
\newblock \textit{ApJ} 640:211

\bibitem[{{Harris} et~al.(2002){Harris}, {Finoguenov}, {Bridle}, {Hardcastle}
  \& {Laing}}]{harr02b}
{Harris} DE, {Finoguenov} A, {Bridle} AH, {Hardcastle} MJ, {Laing} RA. 2002.
\newblock \textit{\apj} 580:110--113

\bibitem[{{Harris} et~al.(1999){Harris}, {Hjorth}, {Sadun}, {Silverman} \&
  {Vestergaard}}]{harr99}
{Harris} DE, {Hjorth} J, {Sadun} AC, {Silverman} JD, {Vestergaard} M. 1999.
\newblock \textit{\apj} 518:213--218

\bibitem[{{Harris} \& {Krawczynski}(2002)}]{harr02a}
{Harris} DE, {Krawczynski} H. 2002.
\newblock \textit{\apj} 565:244--255

\bibitem[{Harris \& Krawczynski(2006)}]{harr06b}
Harris DE, Krawczynski H. 2006.
\newblock In \textit{Triggering Relativstic Jets}, eds. W~Lee, E~Ramirez-Ruiz,
  vol. in press of \textit{Revista Mexicana de Astronomia y Astrofisica, Serie
  de Conferencias}

\bibitem[{{Harris}, {Krawczynski} \& {Taylor}(2002)}]{harr02c}
{Harris} DE, {Krawczynski} H, {Taylor} GB. 2002.
\newblock \textit{\apj} 578:60--63

\bibitem[{Hartman, Bertsch \& Bloom(1999)}]{hart99}
Hartman RC, Bertsch DL, Bloom SD. 1999.
\newblock \textit{\apjs} 123:79

\bibitem[{{Hirotani} et~al.(1999){Hirotani}, {Iguchi}, {Kimura} \&
  {Wajima}}]{hiro99}
{Hirotani} K, {Iguchi} S, {Kimura} M, {Wajima} K. 1999.
\newblock \textit{\pasj} 51:263--267

\bibitem[{{Hughes}(1991)}]{hugh91}
{Hughes} PA. 1991.
\newblock \textit{{Beams and jets in astrophysics}}.
\newblock Press Syndicate of the Univ. of Cambridge, Cambridge University
  Press, Cambridge

\bibitem[{Jester et~al.(2006)Jester, Harris, Marshall, Meisenheimer \&
  Perley}]{jest06}
Jester S, Harris DE, Marshall H, Meisenheimer K, Perley R. 2006.
\newblock \textit{\apj} in press

\bibitem[{{Jester} et~al.(2005){Jester}, {R{\"o}ser}, {Meisenheimer} \&
  {Perley}}]{jest05}
{Jester} S, {R{\"o}ser} HJ, {Meisenheimer} K, {Perley} R. 2005.
\newblock \textit{\aap} 431:477--502

\bibitem[{{Jorstad} \& {Marscher}(2004)}]{jors04a}
{Jorstad} SG, {Marscher} AP. 2004.
\newblock \textit{\apj} 614:615--625

\bibitem[{{Jorstad} et~al.(2005){Jorstad}, {Marscher}, {Lister}, {Stirling},
  {Cawthorne} et~al.}]{jors05}
{Jorstad} SG, {Marscher} AP, {Lister} ML, {Stirling} AM, {Cawthorne} TV, et~al.
  2005.
\newblock \textit{\aj} 130:1418--1465

\bibitem[{Kataoka et~al.(2003)Kataoka, Edwards, Georganopoulos, Takahara \&
  Wagner}]{kata03}
Kataoka J, Edwards P, Georganopoulos M, Takahara F, Wagner S. 2003.
\newblock \textit{A\&A} 399:91

\bibitem[{{Kataoka} \& {Stawarz}(2005)}]{kata05}
{Kataoka} J, {Stawarz} {\L}. 2005.
\newblock \textit{\apj} 622:797--810

\bibitem[{Kataoka et~al.(2006)Kataoka, Stawarz, Aharonian, Takahara, Ostrowski
  \& Edwards}]{kata06}
Kataoka J, Stawarz L, Aharonian F, Takahara F, Ostrowski M, Edwards PG. 2006.
\newblock \textit{\apj} submitted:astro--ph 05100661

\bibitem[{{Kellermann} et~al.(2004){Kellermann}, {Lister}, {Homan},
  {Vermeulen}, {Cohen} et~al.}]{kell04}
{Kellermann} KI, {Lister} ML, {Homan} DC, {Vermeulen} RC, {Cohen} MH, et~al.
  2004.
\newblock \textit{\apj} 609:539--563

\bibitem[{Kino, Takahara \& Kusunose(2002)}]{kino02}
Kino M, Takahara F, Kusunose M. 2002.
\newblock \textit{\apj} 564:97

\bibitem[{Kirk \& Duffy(1999)}]{kirk99}
Kirk JG, Duffy P. 1999.
\newblock \textit{J. Phys. G.} 25:163

\bibitem[{Koide, Shibata \& Kudoh(1999)}]{koid99}
Koide S, Shibata K, Kudoh T. 1999.
\newblock \textit{ApJ} 522:727

\bibitem[{{Komissarov}(1994)}]{komi94}
{Komissarov} SS. 1994.
\newblock \textit{\mnras} 266:649

\bibitem[{K{\"o}nigl(1980)}]{koni80}
K{\"o}nigl A. 1980.
\newblock \textit{Relativistic Effects in Extragalactic Radio Sources}.
\newblock Ph.D. thesis, California Institute of Technology, Pasadena, CA

\bibitem[{{Kraft} et~al.(2002){Kraft}, {Forman}, {Jones}, {Murray},
  {Hardcastle} \& {Worrall}}]{kraf02}
{Kraft} RP, {Forman} WR, {Jones} C, {Murray} SS, {Hardcastle} MJ, {Worrall} DM.
  2002.
\newblock \textit{\apj} 569:54--71

\bibitem[{{Kraft} et~al.(2005){Kraft}, {Hardcastle}, {Worrall} \&
  {Murray}}]{kraf05}
{Kraft} RP, {Hardcastle} MJ, {Worrall} DM, {Murray} SS. 2005.
\newblock \textit{\apj} 622:149--159

\bibitem[{Krawczynski(2004)}]{kraw04}
Krawczynski H. 2004.
\newblock \textit{NewAR} 48:367

\bibitem[{Krawczynski(2005)}]{kraw05}
Krawczynski H. 2005.
\newblock \textit{AIP Proc.} in press, [astro-ph 050862]

\bibitem[{Krawczynski, Coppi \& Aharonian(2002)}]{kraw02}
Krawczynski H, Coppi PS, Aharonian FA. 2002.
\newblock \textit{MNRAS} 336:721

\bibitem[{{Laing} \& {Bridle}(2002{\natexlab{a}})}]{lain02b}
{Laing} RA, {Bridle} AH. 2002{\natexlab{a}}.
\newblock \textit{\mnras} 336:1161--1180

\bibitem[{{Laing} \& {Bridle}(2002{\natexlab{b}})}]{lain02a}
{Laing} RA, {Bridle} AH. 2002{\natexlab{b}}.
\newblock \textit{\mnras} 336:328--352

\bibitem[{{Laing} \& {Bridle}(2004)}]{lain04}
{Laing} RA, {Bridle} AH. 2004.
\newblock \textit{\mnras} 348:1459--1472

\bibitem[{{Laing}, {Canvin} \& {Bridle}(2003)}]{lain03}
{Laing} RA, {Canvin} JR, {Bridle} AH. 2003.
\newblock \textit{New Astronomy Review} 47:577--579

\bibitem[{{Lara} et~al.(2004){Lara}, {Giovannini}, {Cotton}, {Feretti} \&
  {Venturi}}]{lara04}
{Lara} L, {Giovannini} G, {Cotton} WD, {Feretti} L, {Venturi} T. 2004.
\newblock \textit{\aap} 415:905--913

\bibitem[{Lee \& Ramirez-Ruiz(2006)}]{lee06}
Lee WH, Ramirez-Ruiz E, eds. 2006.
\newblock vol. in press. Revista Mexicana de Astronomia y Astrofisica, Serie de
  Conferencias

\bibitem[{{Lister}(2003)}]{list03a}
{Lister} ML. 2003.
\newblock \textit{\apj} 599:105--115

\bibitem[{{Lobanov}, {Hardee} \& {Eilek}(2003)}]{loba03}
{Lobanov} A, {Hardee} P, {Eilek} J. 2003.
\newblock \textit{New Astronomy Review} 47:629--632

\bibitem[{{Lovelace}(1976)}]{love76}
{Lovelace} RVE. 1976.
\newblock \textit{\nat} 262:649--652

\bibitem[{{Ly}, {De Young} \& {Bechtold}(2005)}]{ly05}
{Ly} C, {De Young} DS, {Bechtold} J. 2005.
\newblock \textit{\apj} 618:609--617

\bibitem[{{Macchetto}(1996)}]{macc96}
{Macchetto} FD. 1996.
\newblock In \textit{IAU Symp. 175: Extragalactic Radio Sources}

\bibitem[{{Marshall} et~al.(2001){Marshall}, {Harris}, {Grimes}, {Drake},
  {Fruscione} et~al.}]{mars01}
{Marshall} HL, {Harris} DE, {Grimes} JP, {Drake} JJ, {Fruscione} A, et~al.
  2001.
\newblock \textit{\apjl} 549:L167--L171

\bibitem[{{Marshall} et~al.(2002){Marshall}, {Miller}, {Davis}, {Perlman},
  {Wise} et~al.}]{mars02}
{Marshall} HL, {Miller} BP, {Davis} DS, {Perlman} ES, {Wise} M, et~al. 2002.
\newblock \textit{\apj} 564:683--687

\bibitem[{{Marshall} et~al.(2005){Marshall}, {Schwartz}, {Lovell}, {Murphy},
  {Worrall} et~al.}]{mars05}
{Marshall} HL, {Schwartz} DA, {Lovell} JEJ, {Murphy} DW, {Worrall} DM, et~al.
  2005.
\newblock \textit{\apjs} 156:13--33

\bibitem[{Mattox, Wagner \& Malkan(1997)}]{matt97}
Mattox JR, Wagner SJ, Malkan M. 1997.
\newblock \textit{\apj} 476:692

\bibitem[{McEnery, Moskalenko \& Ormes(2004)}]{mcen04}
McEnery JE, Moskalenko IV, Ormes JF. 2004.
\newblock In \textit{``Cosmic Gamma Ray Sources''}, eds. K~Cheng, G~Romero.
  Kluwer ASSL Series.
\newblock Astro-ph/0406250

\bibitem[{{Nakamura}, {Uchida} \& {Hirose}(2001)}]{naka01}
{Nakamura} M, {Uchida} Y, {Hirose} S. 2001.
\newblock \textit{New Astronomy} 6:61--78

\bibitem[{Nishikawa et~al.(2005)Nishikawa, Hardee, Richardson, Preece, Sol \&
  Fishman}]{nish05}
Nishikawa KI, Hardee P, Richardson G, Preece R, Sol H, Fishman GJ. 2005.
\newblock \textit{\apj} 622:927

\bibitem[{Ostrowski et~al.(1997)Ostrowski, Sikora, Madejski \&
  Begelman}]{ostr97}
Ostrowski M, Sikora M, Madejski G, Begelman M, eds. 1997.
\newblock \textit{Relativistic Jets in AGNs}, Konfederacka 6, 30-306, Krakow.
  Poligrafia Inspecktoratu Towarzystwa Salezjanskiego

\bibitem[{{Perlman} et~al.(2003){Perlman}, {Harris}, {Biretta}, {Sparks} \&
  {Macchetto}}]{perl03}
{Perlman} ES, {Harris} DE, {Biretta} JA, {Sparks} WB, {Macchetto} FD. 2003.
\newblock \textit{\apjl} 599:L65--L68

\bibitem[{{Perlman} \& {Wilson}(2005)}]{perl05}
{Perlman} ES, {Wilson} AS. 2005.
\newblock \textit{\apj} 627:140--155

\bibitem[{{Pesce} et~al.(2001){Pesce}, {Sambruna}, {Tavecchio}, {Maraschi},
  {Cheung} et~al.}]{pesc01}
{Pesce} JE, {Sambruna} RM, {Tavecchio} F, {Maraschi} L, {Cheung} CC, et~al.
  2001.
\newblock \textit{\apjl} 556:L79--L82

\bibitem[{{Pushkarev} et~al.(2005){Pushkarev}, {Gabuzda}, {Vetukhnovskaya} \&
  {Yakimov}}]{push05}
{Pushkarev} AB, {Gabuzda} DC, {Vetukhnovskaya} YN, {Yakimov} VE. 2005.
\newblock \textit{\mnras} 356:859--871

\bibitem[{Rees(1971)}]{rees71}
Rees MJ. 1971.
\newblock \textit{Nature} 229:312

\bibitem[{{Riley} \& {Warner}(1990)}]{rile90}
{Riley} JM, {Warner} PJ. 1990.
\newblock \textit{\mnras} 246:1P

\bibitem[{{Rossi} et~al.(2004){Rossi}, {Bodo}, {Massaglia}, {Ferrari} \&
  {Mignone}}]{ross04}
{Rossi} P, {Bodo} G, {Massaglia} S, {Ferrari} A, {Mignone} A. 2004.
\newblock \textit{\apss} 293:149--155

\bibitem[{{Salpeter}(1964)}]{salp64}
{Salpeter} EE. 1964.
\newblock \textit{\apj} 140:796--800

\bibitem[{Sambruna et~al.(2001)Sambruna, Urry, Tavecchio, Maraschi, Scarpa
  et~al.}]{samb01}
Sambruna R, Urry C, Tavecchio F, Maraschi L, Scarpa R, et~al. 2001.
\newblock \textit{\apj} 549:L161

\bibitem[{{Sambruna} et~al.(2004){Sambruna}, {Gambill}, {Maraschi},
  {Tavecchio}, {Cerutti} et~al.}]{samb04}
{Sambruna} RM, {Gambill} JK, {Maraschi} L, {Tavecchio} F, {Cerutti} R, et~al.
  2004.
\newblock \textit{\apj} 608:698--720

\bibitem[{{Sambruna} et~al.(2002){Sambruna}, {Maraschi}, {Tavecchio}, {Urry},
  {Cheung} et~al.}]{samb02}
{Sambruna} RM, {Maraschi} L, {Tavecchio} F, {Urry} CM, {Cheung} CC, et~al.
  2002.
\newblock \textit{\apj} 571:206--217

\bibitem[{{Sauty}, {Tsinganos} \& {Trussoni}(2002)}]{saut02}
{Sauty} C, {Tsinganos} K, {Trussoni} E. 2002.
\newblock \textit{LNP Vol.~589: Relativistic Flows in Astrophysics} 589:41

\bibitem[{Scheuer(1974)}]{sche74}
Scheuer P. 1974.
\newblock \textit{MNRAS} 166:513

\bibitem[{Schwartz(2002)}]{schw02}
Schwartz DA. 2002.
\newblock \textit{\apj} 569:L23

\bibitem[{Schwartz et~al.(2006)Schwartz, Marshall, Lovell, Murphy, Bicknell
  et~al.}]{schw06}
Schwartz DA, Marshall HL, Lovell JEJ, Murphy DW, Bicknell GV, et~al. 2006.
\newblock \textit{\apj} in press:astro--ph 0601632

\bibitem[{{Schwartz} et~al.(2000){Schwartz}, {Marshall}, {Lovell}, {Piner},
  {Tingay} et~al.}]{schw00}
{Schwartz} DA, {Marshall} HL, {Lovell} JEJ, {Piner} BG, {Tingay} SJ, et~al.
  2000.
\newblock \textit{\apjl} 540:L69

\bibitem[{{Siemiginowska} et~al.(2002){Siemiginowska}, {Bechtold}, {Aldcroft},
  {Elvis}, {Harris} \& {Dobrzycki}}]{siem02}
{Siemiginowska} A, {Bechtold} J, {Aldcroft} TL, {Elvis} M, {Harris} DE,
  {Dobrzycki} A. 2002.
\newblock \textit{\apj} 570:543--556

\bibitem[{{Siemiginowska} et~al.(2003){Siemiginowska}, {Smith}, {Aldcroft},
  {Schwartz}, {Paerels} \& {Petric}}]{siem03b}
{Siemiginowska} A, {Smith} RK, {Aldcroft} TL, {Schwartz} DA, {Paerels} F,
  {Petric} AO. 2003.
\newblock \textit{\apjl} 598:L15--L18

\bibitem[{Siemiginowska et~al.(2003)Siemiginowska, Stanghellini, Brunetti,
  Fiore, Aldcroft et~al.}]{siem03a}
Siemiginowska A, Stanghellini C, Brunetti G, Fiore F, Aldcroft T, et~al. 2003.
\newblock \textit{\apj} 595:643

\bibitem[{{Sikora} et~al.(2005){Sikora}, {Begelman}, {Madejski} \&
  {Lasota}}]{siko05}
{Sikora} M, {Begelman} MC, {Madejski} GM, {Lasota} JP. 2005.
\newblock \textit{\apj} 625:72--77

\bibitem[{Sikora \& Madejski(2001)}]{siko01}
Sikora M, Madejski G. 2001.
\newblock \textit{AIP} 558:275

\bibitem[{{Sikora} et~al.(1997){Sikora}, {Madejski}, {Moderski} \&
  {Poutanen}}]{siko97}
{Sikora} M, {Madejski} G, {Moderski} R, {Poutanen} J. 1997.
\newblock \textit{\apj} 484:108

\bibitem[{{Spada} et~al.(2001){Spada}, {Ghisellini}, {Lazzati} \&
  {Celotti}}]{spad01}
{Spada} M, {Ghisellini} G, {Lazzati} D, {Celotti} A. 2001.
\newblock \textit{\mnras} 325:1559--1570

\bibitem[{{Stawarz}(2004)}]{staw04b}
{Stawarz} {\L}. 2004.
\newblock \textit{\apj} 613:119--128

\bibitem[{{Stawarz} \& {Ostrowski}(2002)}]{staw02}
{Stawarz} {\L}, {Ostrowski} M. 2002.
\newblock \textit{\apj} 578:763--774

\bibitem[{{Stawarz} et~al.(2004){Stawarz}, {Sikora}, {Ostrowski} \&
  {Begelman}}]{staw04a}
{Stawarz} {\L}, {Sikora} M, {Ostrowski} M, {Begelman} MC. 2004.
\newblock \textit{\apj} 608:95--107

\bibitem[{{Swain}, {Bridle} \& {Baum}(1998)}]{swai98}
{Swain} MR, {Bridle} AH, {Baum} SA. 1998.
\newblock \textit{\apjl} 507:L29--L33

\bibitem[{Tanihata et~al.(2003)Tanihata, Takahashi, Kataoka \&
  Madejski}]{tani03}
Tanihata C, Takahashi T, Kataoka J, Madejski GM. 2003.
\newblock \textit{\apj} 584:153

\bibitem[{{Tavecchio}(2004)}]{tave04b}
{Tavecchio} F. 2004.
\newblock \textit{Memorie della Societa Astronomica Italiana Supplement} 5:211

\bibitem[{Tavecchio(2005)}]{tave05}
Tavecchio F. 2005.
\newblock In \textit{Procs. the X'th Marcel Grossmann Meeting on General
  Relativity, Rio de Janeiro, Brazil, July 2003}, vol. astro-ph/0401590

\bibitem[{Tavecchio et~al.(2004)Tavecchio, Maraschi, Sambruna \&
  et~al.}]{tave04a}
Tavecchio F, Maraschi L, Sambruna RM, et~al. 2004.
\newblock \textit{\apj} 614:64

\bibitem[{{Tavecchio} et~al.(2000){Tavecchio}, {Maraschi}, {Sambruna} \&
  {Urry}}]{tave00}
{Tavecchio} F, {Maraschi} L, {Sambruna} RM, {Urry} CM. 2000.
\newblock \textit{\apjl} 544:L23--L26

\bibitem[{{Tingay} et~al.(2000){Tingay}, {Jauncey}, {Reynolds}, {Tzioumis},
  {King} et~al.}]{ting00}
{Tingay} SJ, {Jauncey} DL, {Reynolds} JE, {Tzioumis} AK, {King} EA, et~al.
  2000.
\newblock \textit{Advances in Space Research} 26:677--680

\bibitem[{{Tsinganos} \& {Bogovalov}(2002)}]{tsin02}
{Tsinganos} K, {Bogovalov} S. 2002.
\newblock \textit{\mnras} 337:553--558

\bibitem[{{Urry} \& {Padovani}(1995)}]{urry95}
{Urry} CM, {Padovani} P. 1995.
\newblock \textit{\pasp} 107:803

\bibitem[{Uttley, McHardy \& Vaughan(2005)}]{uttl05}
Uttley P, McHardy IM, Vaughan S. 2005.
\newblock \textit{MNRAS} 359:345

\bibitem[{Walker, Benson \& Unwin(1987)}]{walk87}
Walker R, Benson J, Unwin S. 1987.
\newblock \textit{ApJ} 316:546

\bibitem[{{Wang}(2002)}]{wang02}
{Wang} JC. 2002.
\newblock \textit{Chinese Journal of Astronony and Astrophysics} 2:1--7

\bibitem[{Wardle \& Aaron(1997)}]{ward97}
Wardle JFC, Aaron SE. 1997.
\newblock \textit{MNRAS} 286:425

\bibitem[{{Wardle} et~al.(1998){Wardle}, {Homan}, {Ojha} \& {Roberts}}]{ward98}
{Wardle} JFC, {Homan} DC, {Ojha} R, {Roberts} DH. 1998.
\newblock \textit{\nat} 395:457--461

\bibitem[{Weekes(2003)}]{week03}
Weekes T. 2003.
\newblock \textit{Very high energy gamma-ray astronomy}.
\newblock The Institute of Physics Pub., Bristol, UK

\bibitem[{Weisskopf et~al.(2003)Weisskopf, Aldcroft, Bautz, Cameron, Dewey
  et~al.}]{weis03}
Weisskopf MC, Aldcroft TL, Bautz M, Cameron RA, Dewey D, et~al. 2003.
\newblock \textit{Exper.Astron.} 16:1--68

\bibitem[{Wilson, Young \& Shopbell(2001)}]{wils01}
Wilson A, Young A, Shopbell P. 2001.
\newblock \textit{\apj} 547:740

\bibitem[{Wilson \& Yang(2002)}]{wils02}
Wilson AS, Yang Y. 2002.
\newblock \textit{ApJ} 568:133

\bibitem[{Worrall \& Birkinshaw(2005)}]{worr05}
Worrall D, Birkinshaw M. 2005.
\newblock \textit{MNRAS} 360:926

\bibitem[{Worrall, Birkinshaw \& Hardcastle(2001)}]{worr01}
Worrall D, Birkinshaw M, Hardcastle M. 2001.
\newblock \textit{MNRAS} 326:L7

\bibitem[{{Worrall}, {Birkinshaw} \& {Hardcastle}(2003)}]{worr03}
{Worrall} DM, {Birkinshaw} M, {Hardcastle} MJ. 2003.
\newblock \textit{\mnras} 343:L73--L78

\bibitem[{{Zensus}(1997)}]{zens97}
{Zensus} JA. 1997.
\newblock \textit{Ann. Rev. Ast. \& Astrophys.} 35:607--636

\bibitem[{{Zezas} et~al.(2005){Zezas}, {Birkinshaw}, {Worrall}, {Peters} \&
  {Fabbiano}}]{zeza05}
{Zezas} A, {Birkinshaw} M, {Worrall} DM, {Peters} A, {Fabbiano} G. 2005.
\newblock \textit{\apj} 627:711--720

\end{thebibliography}

\end{document}